\documentclass[journal, 10pt]{IEEEtran}

\usepackage{array}
\usepackage{amsfonts}
\usepackage{amsmath}
\usepackage{latexsym}
\usepackage{graphics}
\usepackage{amsbsy}
\usepackage{amssymb}
\usepackage{algorithm}
\usepackage{algorithmic}
\usepackage{enumerate}
\usepackage{color}
\usepackage{cite}
\usepackage{verbatim}
\usepackage{graphicx}%
\usepackage{multirow}
\usepackage{pifont}
\usepackage[usenames,dvipsnames]{xcolor}
\usepackage{algorithm}
\usepackage{array}
\usepackage{amsfonts}
\usepackage{amsmath}
\usepackage{latexsym}
\usepackage{graphics}
\usepackage{amsbsy}
\usepackage{amssymb}
\usepackage{algorithm}
\usepackage{algorithmic}
\usepackage{enumerate}
\usepackage{color}
\usepackage{cite}
\usepackage{verbatim}
\usepackage{graphicx}%
\usepackage{multirow}
\usepackage{pifont}
\usepackage[usenames,dvipsnames]{xcolor}
\usepackage{algorithm}

\begin{document}

\title{{\color{black} Blockchain as a Service: A Decentralized and Secure Computing Paradigm} 
}
\author{
    \IEEEauthorblockN{Gihan J. Mendis\IEEEauthorrefmark{1}, Yifu Wu\IEEEauthorrefmark{1}, Jin Wei\IEEEauthorrefmark{1}, Moein Sabounchi\IEEEauthorrefmark{1}, and Rigoberto Roche'\IEEEauthorrefmark{2}}\\
    \IEEEauthorblockA{\IEEEauthorrefmark{1}Department of Computer and Information Technology\\
	Purdue University, West Lafayette, Indiana}\\
    \IEEEauthorblockA{\IEEEauthorrefmark{2}NASA Glenn Research Center, Cleveland, Ohio}
}
%
%
%


\maketitle
\begin{abstract}
Thanks to the advances in machine learning, data-driven analysis tools have become valuable solutions for various applications. However there still remain essential challenges to develop effective data-driven methods because of the need to acquire a large amount of data and to have sufficient computing power to handle the data. In many instances these challenges are addressed by relying on a dominant cloud computing vendor, but, although commercial cloud vendors provide valuable platforms for data analytics, they can suffer from a lack of transparency, security, and privacy-perservation. Furthermore, reliance on cloud servers prevents applying big data analytics in environments where the computing power is scattered. To address these challenges, a decentralize, secure, and privacy-preserving computing paradigm is proposed to enable an asynchronized cooperative computing process amongst scattered and untrustworthy computing nodes that may have limited computing power and computing intelligence. This paradigm is designed by exploring blockchain, decentralized learning, homomorphic encryption, and software defined networking(SDN) techniques. The performance of the proposed paradigm is evaluated via different scenarios in the simulation section.

\end{abstract}


\section{Introduction}

Due to the advances in sensing and computing, data-driven methods, such as machine learning techniques~\cite{van2018artificial}, have become very promising solutions for different applications~\cite{lecun2015deepNature,goodfellow2016deep,schmidhuber2015deep}. However, there are two essential challenges in implementing these techniques: (1) acquisition of large amount of data, and (2) requirement of enough computing power, which enforces the reliance on a dominant cloud computing vendor. Although the cloud servers provide valuable platforms for big data analytics, there remain several essential challenges in adopting the commercial cloud vendors: (1) transparency, (2) security, and (3) privacy~\cite{Pan2012}. With the rise of awareness for data privacy, end-users have become reluctant to share their data. Also, in another perspective, user data are becoming a valuable asset. In some domains, such as the healthcare sector, federal and civil service offices, there is an abundance of valuable data, however, due to privacy laws and regulations, these data cannot be shared with the third party. Furthermore, the reliance on cloud servers also limit the potentials of applying big data analytics for the environment where the computing power is scattered. Therefore, it is meaningful to develop a reliable infrastructure in which end-users or data creators are able to secure the ownership of data while being able to contribute to machine learning tasks in a privacy-preserving manner with reasonable financial incentivation. To achieve this goal, in this paper we develop a decentralized and secure computing infrastructure that enables an effective and privacy-preserving collaboration between the available end-users and data creators that are called computing nodes in this paper. These computing nodes can have restricted computing power and limited computing intelligence. Additionally, they can be scattered and untrustworthy to each other.

In recent years, several techniques have been proposed to achieve decentralized and privacy-preserving computing. In~\cite{shokri2015privacy}, Shokri \textit{et al.} proposed a privacy-preserving deep-learning mechanism with secure multi-party computations to update the single initial deep learning model. However, the deep learning model is centralized while multiple parties contribute to the model training in a manner that guaranties the privacy-preservation. Federated learning introduced in~\cite{konevcny2016federated,konevcny2016federated2,Fedarate2017,BonEichGrie19} is a distributed machine learning method that enables model training on decentralized data. In this method, multiple copies of the central model can be available in distributed computing devices for training, which eliminates the single point of failure. However, both of these two methods must be managed by one centralized authoritative controlling agent, which may raise security and privacy concerns. To address these issues, in our work we develop a decentralized and secure computing paradigm, which does not have any centralized authoritative controlling agents, by exploiting blockchain, machine learning, and homomorphic encryption technologies. Blockchain is an emerging technology, which can be considered as an immutable and decentralized digital ledger~\cite{Ethereum13,Ethereum16}. A blockchain is a growing list of records, called blocks, which are linked via cryptography. Each block contains the hash value of the previous block, the transaction data, and the timestamp. Due to the inherent cryptographic chaining, if a malicious party tries to manipulate certain transaction data, it will causes the changes of the hash values of the block containing the transaction and those of all the subsequent blocks, which can be easily detected. Therefore, generally speaking, blockchain technology provides a very promising solution for integrity security. Amongst various existing blockchain platforms, Ethereum and Bitcoin are two of the most widely adopted ones~\cite{Bitcoin09,Ethereum13,Ethereum16}. Compared with Bitcoin, Ethereum platform provides a trustful automation of programs via smart contracts that run on virtual machines. In our work, we exploit Ethereum blockchain to execute the secure and privacy-preserving decentralized computing functionalities automatically.

Furthermore, our computing paradigm enables the effective decentralized and cooperative learning via an effective learning-model fusion mechanism. Fusing multiple learning models is an active area of research and fusion strategies in literature can be divided into two main categories: (1) \emph{Late fusion} that comprise predicting the labels based on the labels given by each learning model to be fused and (2) \emph{Early fusion} that takes the feature vectors given by the individual learning models as the inputs and learns a classifier on top of them. Although \emph{late fusion} requires lower computational cost compared with \emph{early fusion} in many practical applications as stated in~\cite{vielzeuf2017temporal,kahou2013combining,ye2012robust}, \emph{early fusion} can achieve a more optimal way to combine learned models compared with \emph{late fusion}~\cite{neverova2016moddrop}. In this work, our learning-model fusion mechanism belongs to the type of \emph{early fusion}, in which the feature vectors to be fused present features with the highest level of abstraction. Additionally, we consider two strategies for designing the fusion mechanism: (1) using a fully connected structure with a single hidden-layer to map concatenated features to labels and (2) implementing gradual fusion to explore the uniqueness of the individual learning models and the correlation amongst the learning models.

To further enhance the security and achieve privacy-preservation, we design a encryption interface with a zero-knowledge proof protocol by exploiting homomorphic encryption (HE), which enables evaluating the performance of the contributed learning models without revealing the sensitive details of the learning models. HE technology is one form of encryption that allows the computation operations to be directly implemented in cipherspace and achieves an encrypted results that, when decrypted, match the results of the operations as if they had been performed on the plainspace~\cite{rivest1978data}. The existing HE technologies can be generally classified into three main groups: (1) fully HE schemes, (2) partially HE schemes, and (3) somewhat HE schemes~\cite{rivest1978method,HE14,gentry2009fully,FullyHE15,elgamal1985public,zhou2014efficient,HEIV15}. Considering the fact that somewhat HE schemes support more operations compared with partially HE schemes and require less computation power compared with fully HE schemes, we exploit Integer-Vector HE scheme~\cite{zhou2014efficient,HEIV15}, which is a somewhat HE scheme, to develop the encryption interface in our computing paradigm. {\color{black}The authors would like to claim that the technologies presented in this paper has been include in a provisional patent~\cite{patent}.}

The next section describes the problem setting for our work. Section~\ref{sec:proposed} describes our proposed blockchain-powered decentralized and secure computing paradigm  mechanism followed by the details of the implementation in Section~\ref{sec:implementation}. Simulation results and the conclusions are shown in Sections~\ref{sec:sim}~and~\ref{sec:con}, repectively.

\section{Problem Setting}
 \begin{figure*}[]
\center
\includegraphics[width=1\textwidth]{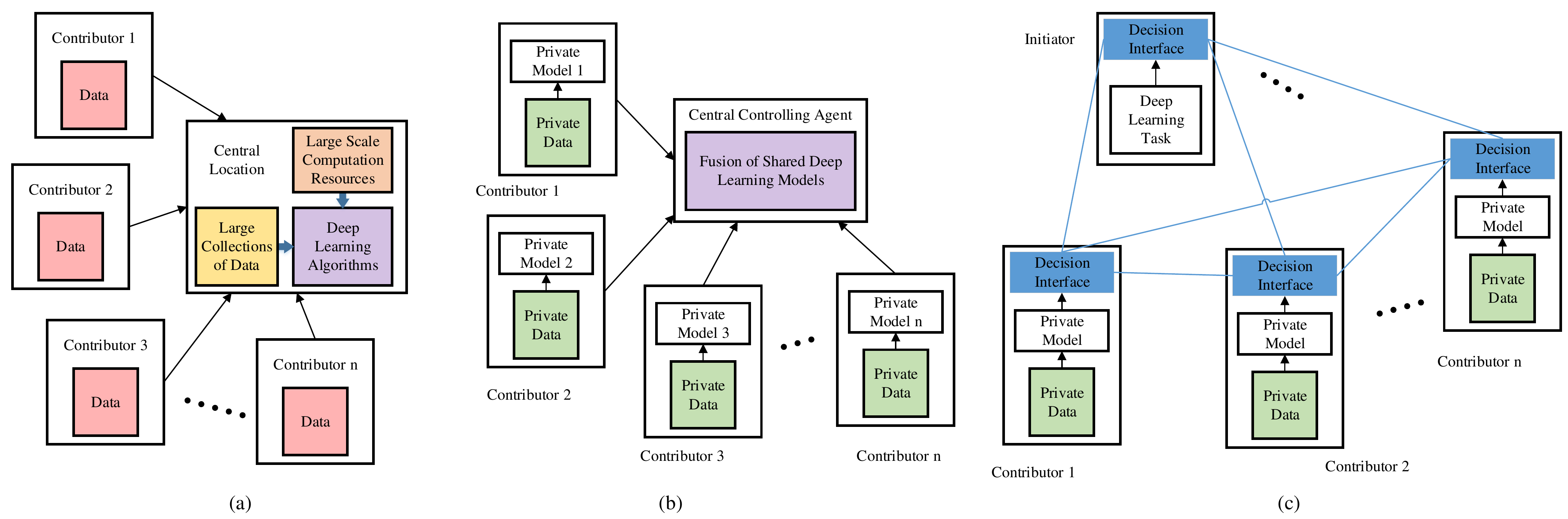}
\caption{\label{Fig:central} (a) Centralized machine learning architecture where data are collected to centralized server with high processing and storage capability; (b) Distributed machine learning architecture where partial of training is distributed to the data contributors and the training process is fully controlled by a central controlling agent;
(c) Autonomous cooperative and decentralized machine learning architecture with no central agents facilitated by blockchain service infrastructure.}
\center
\end{figure*}
Two of the main factors for the thriving of machine learning are: (1) the availability of sizable data-sets with generalized distributions, commonly known as big data, and (2) the availability of the computational power to process this big data, mainly in the form of large-scale GPU clusters. Because of this, most profitable parties in the field of machine learning are large organizations, which hold both valuable big data and sufficient computational power to process it. As illustrated in Fig.~\ref{Fig:central}(a), these large organizations collect data from data contributors to advance the capabilities of their machine learning techniques. One of the essential challenges for creating large-scale datasets via data acquisition, from multiple parties, is the issue of privacy and the related concerns. Potential data providers may not get motivated to share the data because of the high potential for data privacy violations. Additionally, collecting tremendous raw data from multiple parties results in a huge demand on communication bandwidth and a dramatically increased attack surface. Furthermore, a large amount of computational power is required by the central server to process the collected big data.

One solution is the implementation of distributed learning architectures~\cite{shokri2015privacy,konevcny2016federated,konevcny2016federated2,Fedarate2017,BonEichGrie19}. As shown in Fig.~\ref{Fig:central}(b), in distributed learning, rather than collecting and processing data in a single central server, data processing is distributed partially to the individual data providers. By doing so, the distributed learning is implemented in such a way, that the computing contributors process their local data by training the given machine learning models or their own machine learning models and then share the trained model with a central controlling agent. Since the data are not shared, we can say that the data privacy is preserved in this architecture. Additionally, the machine learning models are trained in distributed locations with smaller sets of data, and thus the computational power required by the individual computing contributors is much lower, compared with that of a central server. However, in this solution, the machine learning architecture is fully controlled by a authoritative agent in a centralized manner. It relies on the central authority to coordinate the activities of each entity in the system. Therefore, it is required that the computing contributors trust the central controlling agent, which may raise security and privacy concerns.

To mitigate this, we improve the distributed machine learning architecture presented in Fig.~\ref{Fig:central}(b) and achieve the decentralized and cooperative machine learning architecture shown in Fig.~\ref{Fig:central}(c). In a decentralized system, each entity is completely autonomous and responsible for its own individual behavior. In this architecture, the untrustworthy computing contributors have full control on their own deep learning models and private data. Additionally, the individual contributors are able to participate or leave the computing architecture, without disturbing the functionality and efficiency of the overall learning process. Also, the participation of the computing contributors is motivated by financial compensation that they will receive according to the value of their contribution. To achieve these objectives, we exploit the Ethereum blockchain and design the smart contract to secure the peer-to-peer transactions between the multiple untrustworthy parties to enable the autonomous decentralized and cooperative deep learning.

\section{Proposed Blockchain-Empowered Cooperative Machine Learning Platform}\label{sec:proposed}

 \begin{figure}[]
\center
\includegraphics[width=0.5\textwidth]{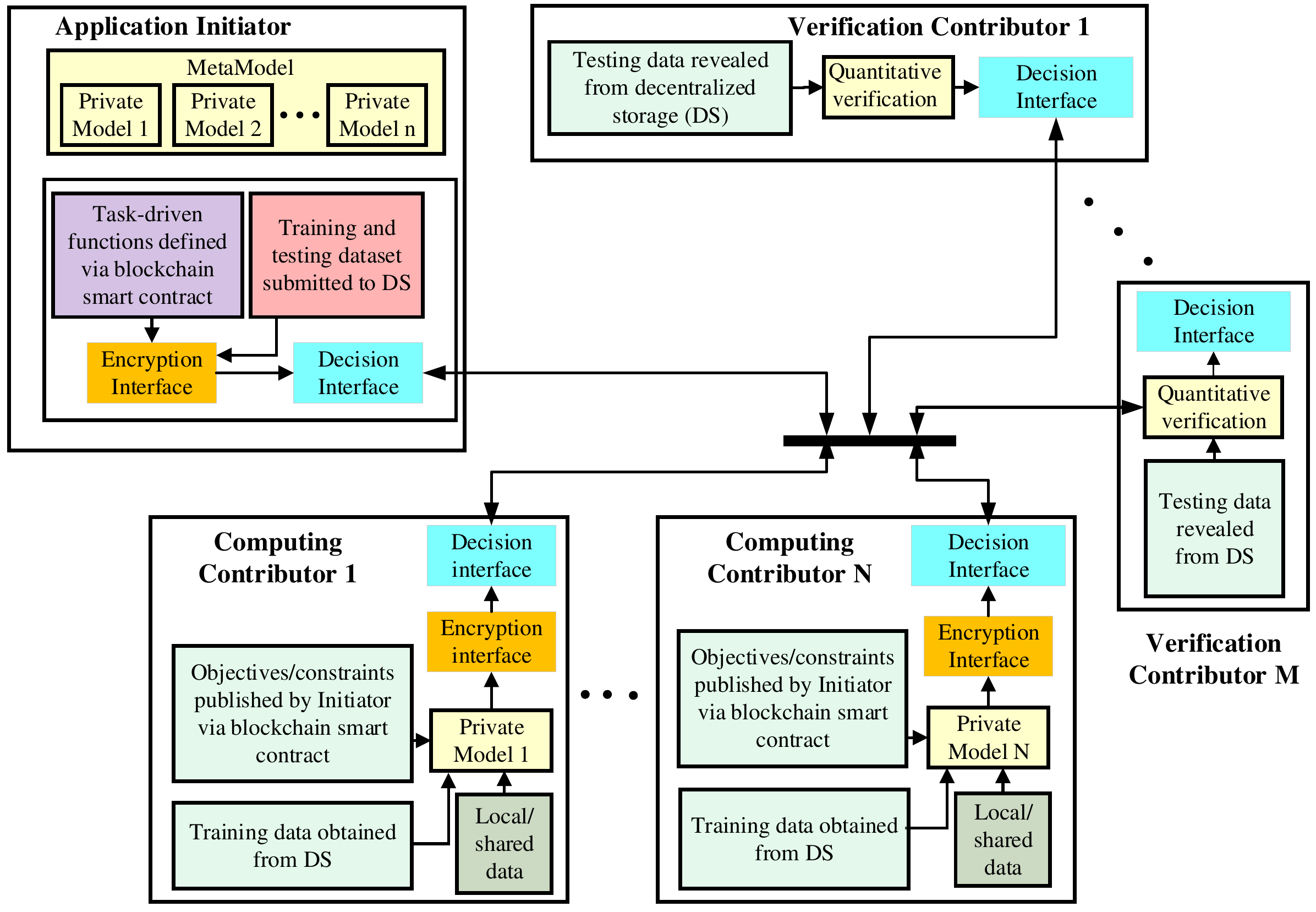}
\caption{\label{Fig:DL}Overview of our blockchain-powered decentralized and secure computing paradigm.}
\center
\end{figure}

\begin{figure}[]
\center
\includegraphics[width=0.5\textwidth]{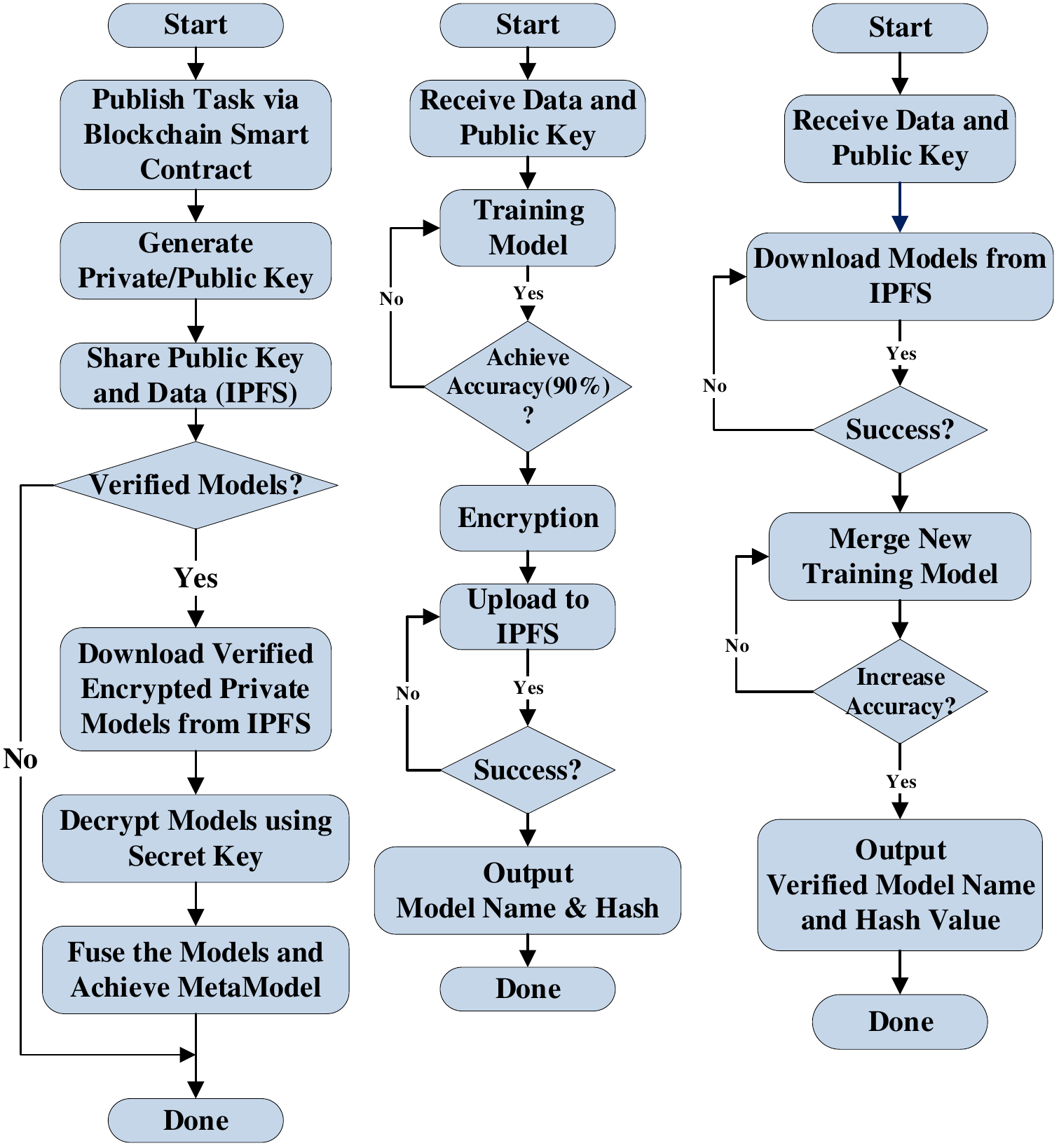}\\
~~~~(a)~~~~~~~~~~~~~~~~~~(b)~~~~~~~~~~~~~~~~~~~~~(c)
\caption{\label{Fig:DL_1}The schematic diagram showing workflows for (a) application initiators, (b) computing contributors, and (c) verification contributors.
}
\center
\end{figure}

The overview of our proposed blockchain-powered decentralized and decentralized computing paradigm is illustrated in Fig.~\ref{Fig:DL}. As shown in Fig.~\ref{Fig:DL}, our proposed mechanism is designed to enable the effective cooperation between the available and possibly scattered computing nodes to accomplish data-driven task that may require high computing power and intelligent and large dataset. The individual computing nodes can participate the computing paradigm by playing one of the three roles: (1) application initiators, (2) computing contributors, or (3) verification contributors. As detailed in Fig.~\ref{Fig:DL_1}(a), if the computing node act as an application initiator, it announces the data-driven applications and is responsible for defining the computing tasks, such as the objectives, constraints, the suggestions on the computing model structure, and the financial compensation commitments, via blockchain smart contract. The application initiators also provide the verification contributors with a sample set of data to evaluate the performance of the learning models contributed by the computing contributors. If it is necessary, the application contributors also provide the computing contributors, which have computing power and computing intelligence, with the dataset to conduct the local training. The sharing of the dataset is realized via the decentralized storage (DS) such as The Interplanetary File System (IPFS)~\cite{IPFS17}. As shown in Fig.~\ref{Fig:DL_1}(b), the computing contributors train the machine-learning models locally for a given data-driven task by using a certain local data asset or the data shared by the associated application initiator. After training the local learning model successfully according to the criteria defined by the application initiator via smart contract, such as the accuracy is above $90$~\%, the computing contributors announce the completeness of the training via the blockchain platform and share the achieved the machine-learning model to the randomly selected verification contributors via the DS such as IPFS. The available verification contributors are passively and randomly selected to provide the hardware resources and verify the contributions of the locally trained learning models, which are claimed by the computing contributors, in a random and decentralized manner. As illustrated in Fig.~\ref{Fig:DL_1}(c), the verification is conducted according to the criteria defined by the application initiator via smart contract, such as whether the accuracy can be improved after fusing the claimed model. The majority voting amongst the verification contributors is used to determine the contribution of the corresponding computing contributors. The application initiator is informed about the majority voting conclusion. If this conclusion is positive, the transaction of the verified locally-trained machine-learning model, also called private model, is established between the application initiator and the associated computing contributor, in which the computing contributor receives the financial compensation and the application initiator obtains the access to the learning model in IPFS. Additionally, the verification contributors involved in the task also get compensated from initiators for their effort. After the time window assigned to the data-driven task ends, the application initiator fuses all the received verified private models to achieve the MetaModel that will be applied to address the data-driven task. The architecture of our proposed computing paradigm is considered asynchronous and adaptive since the computing and verification contributors can leave or join the task at their convenience.

\section{Implementation of Proposed Computing Paradigm}\label{sec:implementation}
In this section, we described the implementation of our proposed blockchain-powered decentralized and secure computing paradigm in details. The implementation of our proposed computing paradigm comprises of three layers: (1) application layer to conduct decentralized and secure computing, also called computing layer, (2) blockchain middleware to enable the secure peer-to-peer communications between the computing nodes including the transactions of the private models and the data sharing, and (3) software-defined networking (SDN)-enabled networking layer.
\subsection{Decentralized Storage System}
As shown in Fig.~\ref{Fig:DL}, decentralized storage (DS) system is one essential component of the computing layer to secure the sharing of data and private machine-learning models. DS system makes the access these data and learning models more affordable, faster and safer. Firstly, expensive centralized data servers are no longer needed in a DS system because every peer can cache the data required by other peers. Secondly, optimized downloading from multiple local peers provides higher traffic throughput than that from a remote centralized data server. In addition, as shown in Fig.~\ref{Fig:DL_1}, after the data or private models are stored in the DS system, the unique hashes characterizing the fingerprint of these cyber assets are generated and shared, which ensures the integrity and the authorization of the sharing process.

\subsection{Blockchain Middleware}
Ethereum Blockchain-based middleware is designed to automatically control, manage and secure the system processes. First of all, consensus protocol, such as Proof-of-Work (PoW), Proof-of-Stake (PoS), and Proof-of-Authority (PoA) of blockchain provides an unsupervised secure environment where the authoritative agents are not necessary any more for the decentralized computing. Additionally, the distributed database of blockchain provides a shared, irremovable ledge of any events happened in time order on the system. It is convenient to trace the ins and outs of a event on blockchain ledge. Furthermore, blockchain smart contract enables the automation of system processes including the training, verification, transaction, and fusion processes of the decentralized and cooperative learning.

\subsection{Homomorphic Encryption Interface}\label{sec:intvect1}

\begin{figure}[]
\center
\includegraphics[width=0.5\textwidth]{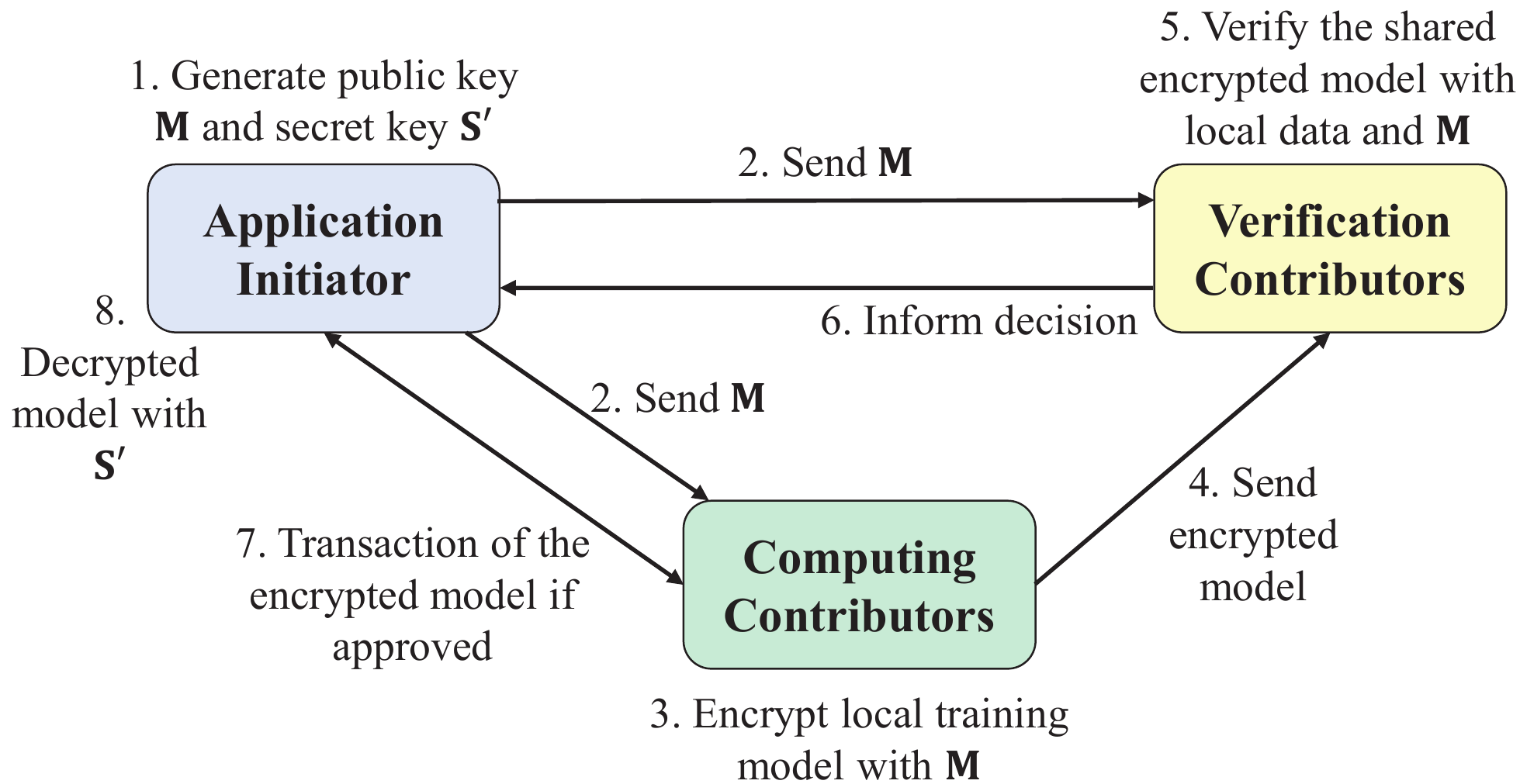}
\caption{The illustration of the mechanism of our encryption interface.}\label{Fig:Encryption}
\end{figure}

As shown in Fig.~\ref{Fig:Encryption}, the encryption interface is designed to enhance the security and enable the privacy-preservation of our computing paradigm. In this work, the encryption interface is developed by exploiting Integer Vector Homomorphic Encryption (HE) scheme. Figure~\ref{Fig:Encryption} illustrates the overall mechanism of the encryption interface, which mainly consist of eight steps. Step 1: Encryption interface client of the application initiator generates the public key $\bf{M}$ and the secret key $\bf{S'}$ according to Integer Vector HE scheme, which is illustrated in Fig.~\ref{Fig:DL_1}(a). Step 2: The generated public key $\bf{M}$ is shared amongst the active computing and verification contributors, which is illustrated in Figs.~\ref{Fig:DL_1}(a)-(c). Steps 3 and 4: Computing contributors apply the received public key $\bf{M}$ to encrypt the machine learning models achieved locally and share the encrypted private models with the passively selected verification contributors for quantitative verification, which is illustrated in Fig.~\ref{Fig:DL_1}(b). Step 5: After receiving the encrypted private model, the verification contributors verify the performance of the private models by conducting the quantitative verification in cipherspace with the public key $\bf{M}$. In the quantitative verification, the verification contributor fuses the received private model with the existing learning model in cipherspace and concludes that the private model is valuable if the overall accuracy increases after model fusion. Step 6: The majority voting amongst all the associated verification contributors is used to determine the contribution of the private model. The application initiator is informed about the majority voting conclusion. Step 7: If the majority conclusion is positive, the transaction of the encrypted private model between the associated computing contributor and the application initiator is established. At last, the application initiator decrypts the shared model with the secret key $\bf{S'}$.

\subsubsection{Integer-Vector Homomorphic Encryption}\label{sec:intvect}

In our work, the integer-vector homomorphic encryption scheme, which supports three fundamental operations: addition, linear transformation, and weighted inner products, is exploited to develop the encryption interface. Letting $x\in \mathbb{Z}^{m}$ be the plaintext vector, $\bf{S} \in \mathbb{Z}^{m \times n}$ be the secret key, and $c \in \mathbb{Z}^{n}$ be the ciphertext vector, the encryption can be formulated as:

\begin{equation}
Sc = wx + e
\end{equation}where$e$ is a randomization term introduced to enable the encryption, which have elements smaller than $w$, and  $w$ is a large integer that controls the appropriate ratio between the plaintext and the introduced randomization.

Given the secret key $S$, the decryption can be performed as:
\begin{equation}
x = \lceil \dfrac{Sc}{w} \rfloor
\end{equation}

In this homomorphic encryption scheme, \emph{key switching} method was proposed to convert the ciphertext in one cipherspace to another without decryption. This convert is realized via a public key $\bf{M}$ that is calculated as follows:

\begin{equation}\label{eq:public}
\bf{M} = \begin{bmatrix}
\bf{S^*}-\bf{TA}+\bf{E} \\
\bf{A}
\end{bmatrix}
\end{equation} where $\bf{S}$ and $\bf{c}$ are the original secret key and ciphertext, respectively, $\bf{S^*}$ is an intermediate key satisfying $\bf{S^{*}c^{*}}=\bf{Sc}$,  $\bf{c^*}$ is the representation of $\bf{c}$ with each digit of $\bf{c}$, $c_i$ represented as a $l$-bit binary number, and $\bf{A}$ and $\bf{E}$ are random and bounded matrices, respectively. A scalar $l$ is selected to be large enough such that $|\mathbf{c}| < 2^l$, which determines the maximum value of $\bf{c}$ represented with $l$ bits. Let $\bf{b}_i$ be the bit representation of the value $c_i$, we can obtain $\bf{c^*}$ as follows,
\begin{equation}
\bf{c^*} =\left[\bf{b}_i,\bf{b}_{i+1},..., \bf{b}_n\right]
\end{equation} where $n$ is the length of vector $\bf{c}$.
Similarly, $\bf{S^*}$ can be obtained as follows:
\begin{equation}
\mathbf{B}_{ij} =\left[2^{l-1}S_{ij},..., 2 S_{ij}, S_{ij}\right]
\end{equation}where $\bf{B}_{ij}$ is the sub-vector of $\bf{S^*}$ that corresponds to the element $S_{ij}$.

Additionally, since the  initial secret key $\bf{S}$ is a identity matrix of the dimension $n \times n$ , the original ciphertext $\bf{c}$ is actually the original plaintext $\bf{x}$ itself. Let $\bf{S'} = [\bf{I}, \bf{T}]$, where $\bf{I}$ is identity matrix and $\bf{T}$, is a desired secret vector. By using the public key $\bf{M}$ defined in Eq.~(\ref{eq:public}), the ciphertext corresponding to the desired secret key $\bf{S'}$ can be calculated as: \begin{equation}
\bf{c'} = \bf{M}c^*
\end{equation}where $\bf{M}$ is a $(n+1)\times nl$ dimension matrix. Therefore, the resulting ciphertext $\bf{c'}$ is a integer vector with length $n+1$.

\subsubsection{Implementation of Artificial Neural Networks-based Machine Learning Model in Cipherspace}

As shown in Fig.~\ref{Fig:Encryption}, the essential component of our homomorphic encryption is to implement the artificial neural network (ANN)-based machine learning model in cipherspace by using integer-vector homomorphic encryption (IVHE) scheme. ANN implementation mainly comprises of summation, vector addition, vector multiplication with scalar, vector dot product, pooling operations, convolution operations, and nonlinear activation functions. Most of these operations are supported by the IVHE scheme, except pooling operations, convolution operations, and non-linear activation function realizations. To enable the implementation of pooling operations in cipherspace, we currently assume the computing contributors adopt the average pooling or summation pooling for training their machine learning model locally. Under this assumption, the pool operations can be realized via the summation followed  by a division by a integer, which is supported by the IVHE. The convolution operations can be implemented in cipherspace by calculating the vector multiplication operations. Additionally, we mainly consider two types of activation functions in the current work: sigmoid function, $\sigma(x) = 1/(1+e^{-x})$, and ReLU function, $ReLU(x) = max(0, x)$. To implement the sigmoid functions in cipherspace, we leverage the Taylor series expansion to achieve the $k$-th order polynomial approximation of the sigmoid function. For example, if $k=3$, the polynomial approximation is $\bar{\sigma}= \dfrac{1}{2} + \dfrac{x}{4} - \dfrac{x^3}{348}$ that is supported by the IVHE scheme. To enable the implementation of ReLU function, which is a piecewise linear operation with discontinuity at $x=0$, in cipherspace, we currently constrain the secret and public keys to contain the non-negative elements only. By doing so, there is no sign changes while encrypting. The ReLU function is executed in cipherspace via the random dropout.

There still remain two challenges of implementing the ANN-based machine learning model in cipherspace. First, most weight values and input data for ANNs are floating-point values that cannot directly be supported by our adopted homomorphic encryption scheme. Additionally, implementing the average pooling and calculating the polynomial approximation of sigmoid function require the multiplications with floating-point numbers, which is also not supported by our adopted homomorphic encryption scheme. To address this issue, we introduce predetermined and unified scaling factors to convert the floating-poitn values to integers. A final propagated scaling factor is used to scale down the output values of the ANNs to the original values. Second, as discussed in Section~\ref{sec:intvect}, inter-vector homomorphic encryption increases the length of ciphertext vector by $1$ compared with that of the plaintext vector. This difference between the dimensions of the ciphertext and plaintext raises challenges in implementing the ANN operations that requires consistency in dimensions such as feed-forward operations. To address this issue, we develop two encryption strategies as follows:

\paragraph{Element-wise Encryption Strategy}\label{sec:hom1}

\begin{figure*}[]
\center
\includegraphics[width=0.4\textwidth]{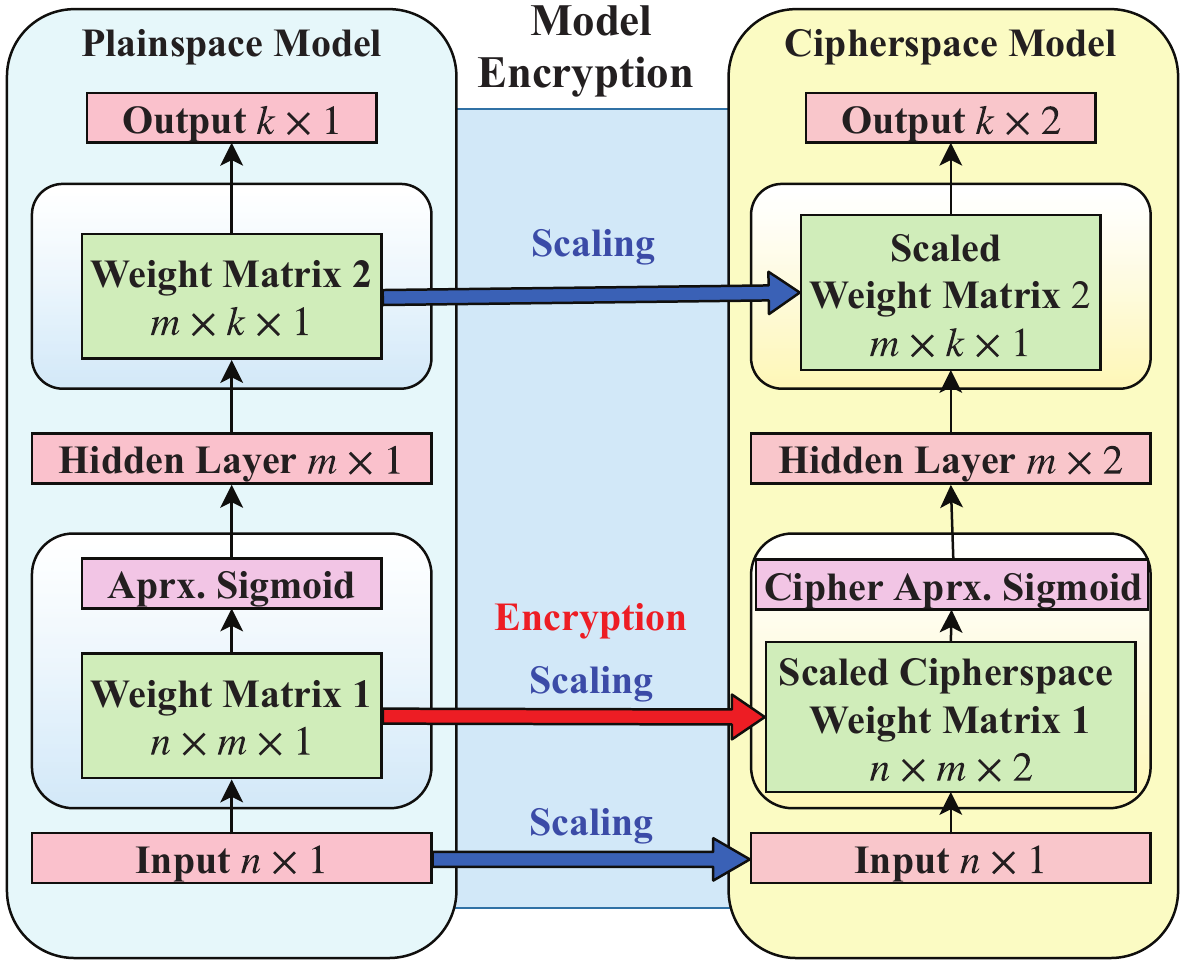}~
\includegraphics[width=0.2\textwidth]{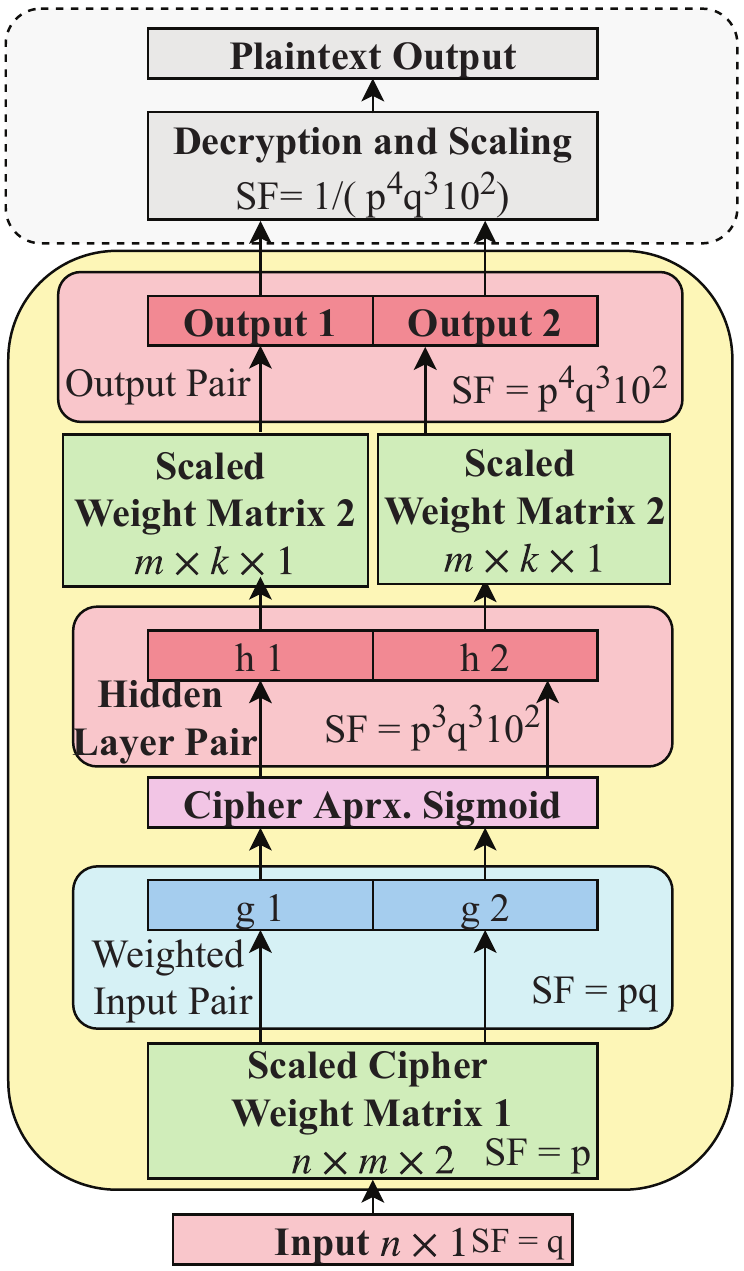}~
\includegraphics[width=0.2\textwidth]{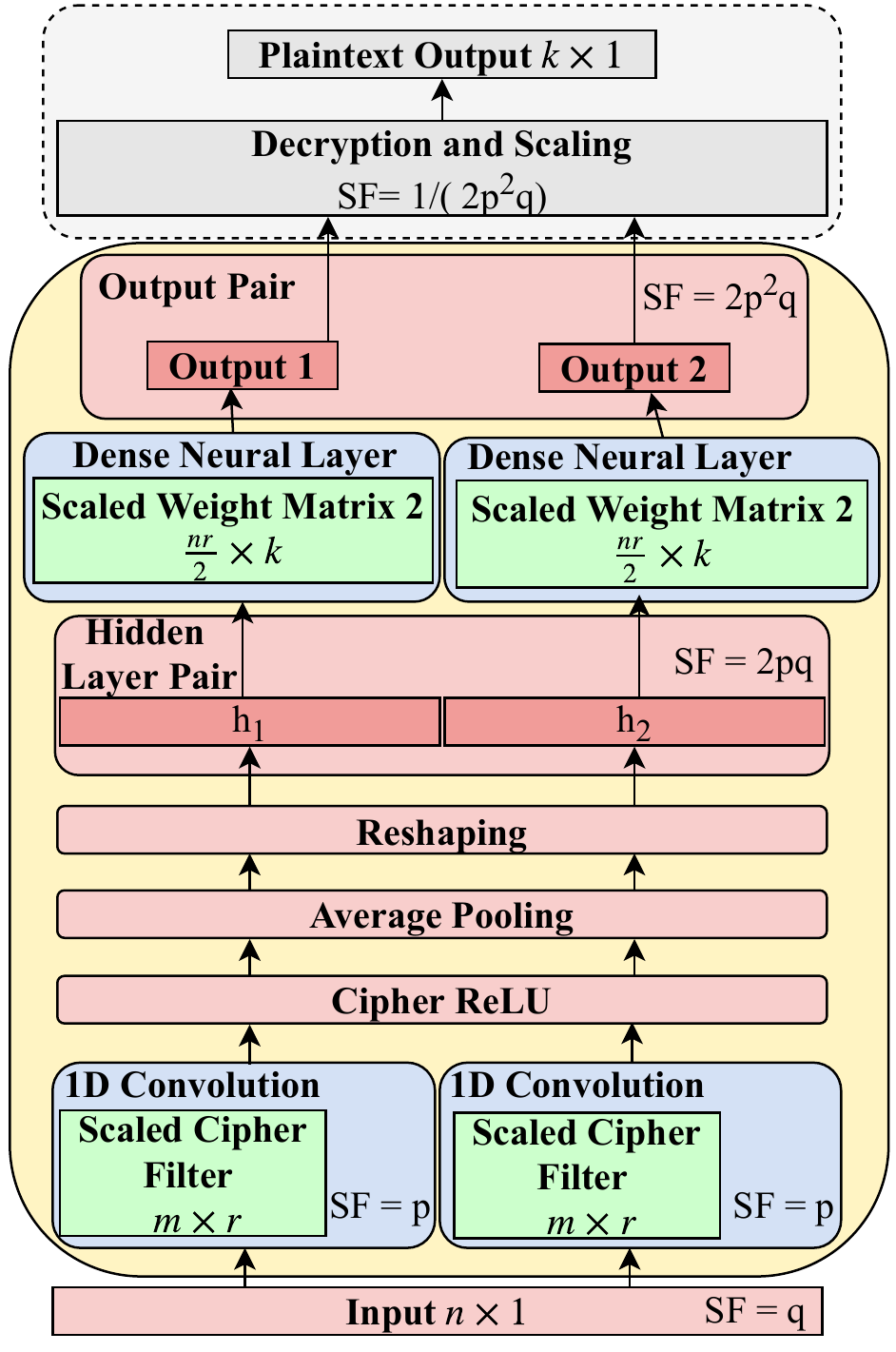}\\
~~~~~~~~~~~(a)~~~~~~~~~~~~~~~~~~~~~~~~~~~~~~~~~~~~~~~~~~~~~~(b)~~~~~~~~~~~~~~~~~~~~~~~~~~~~~(c)
\caption{The illustration of our element-wise encryption strategy: (a) using a fully-connected NN as an example, (b) detailing the execution of the fully-connected NN in cipherspace, and (c) using a CNN as an example.}\label{Fig:Method-1}
\end{figure*}

The essential idea of our element-wise encryption strategy is to encrypt the matrices and vectors in a manner of element by element. By doing so, the additional components introduced by the homomorphic encryption can be addressed in the third dimension, which ensures the consistency on the original dimensions. To illustrate our strategy, we use a fully-connected neural network (NN) as an example shown in Fig.~\ref{Fig:Method-1}(a). The details of the implementation of the fully-connected NN in cipherspace are illustrated in Fig.~\ref{Fig:Method-1}(b).

As shown in Fig~\ref{Fig:Method-1}(a), the fully-connected ANN has an input vectors with length of $n$, a weight matrix, Weight Matrix 1, with the dimension of $n \times m$ resulting in a hidden layer of size $m$ adopting sigmoid function as the activation function, a weight matrix, Weight Matrix 2, with dimension $m \times k$, and a output vector of length $k$. To enable the implantation of our element-wise encryption strategy, Weight Matrices 1 and 2 are represented with dimensions $n \times m \times 1$ and $n \times k \times 1$, respectively, and multiplied with a scaling factor $p$ to ensure all the matrix elements to be integer. Another scaling factor $q$ is introduced to convert the input elements to be integer. Considering the fact that, if Weight Matrix is encrypted, the following structure of the fully-connected NN is meaningless for a malicious party, it is reasonable to focus on encrypting Weight Matrix 1 only to achieve a tradeoff between the high security and low computational complexity. As shown in Figs.~\ref{Fig:Method-1}(a)~and~(b), we consider each element of Weight Matrix 1 as a vector with length $1$ in the 3rd dimension, and apply the element-wise encryption strategy, which results in a weight matrix with the dimension of $n \times m \times 2$ in cipherspace. The dot product operation between the scaled and encrypted Weight Matrix 1 and the scaled input vectors is executed by using IVHE, which results in a pair of weighted inputs in cipherspace. By achieving a polynomial approximation of the sigmoid activation function, the hidden layer with the approximated sigmoid functions is encrypted via the homomorphic encryption resulting in a hidden layer pair in cipherspace. For each of the encrypted hidden layer in the pair, a dot product operation is performed with the scaled Weight Matrix 2, which results in an output pair in cipherspace. Additionally, as shown in Fig~\ref{Fig:Method-1}(b), a scaling-down operation is required during the decryption conducted in the encryption interface in application initiator.

Figure~\ref{Fig:Method-1}(c) illustrates the implementation of our element-wise encryption strategy to execute a convolutional neural network (CNN). Similar to the previous example, the convolution filter is converted to a pair of scaled and encrypted filters having the dimension of $m \times r$. A dot product operation is performed on each encrypted convolution filter of the pair with the scaled input vector. Then a pair of hidden layers in cipherspace is achieved by implementing ReLU operations, sum-pooling, and reshaping. Another dot product operation is executed on each of the hidden layer in the pair with the scaled Weight Matrix 2, which results in an output pair in cipherspace. Since, in this example structure, the encrypted ReLU activation function is realized via random dropout and the operations of sum-pooling and reshaping are executed in cipherspace using IVHE scheme, no additional scaling is introduced through these operations. Therefore, the final output scaling factor remains lower compared to the previous example fully-connected NN using sigmoid activation function.

\paragraph{Matrix-Pair-wise Encryption Strategy}\label{sec:hom2}

\begin{figure}[]
\center
\includegraphics[width=0.3\textwidth]{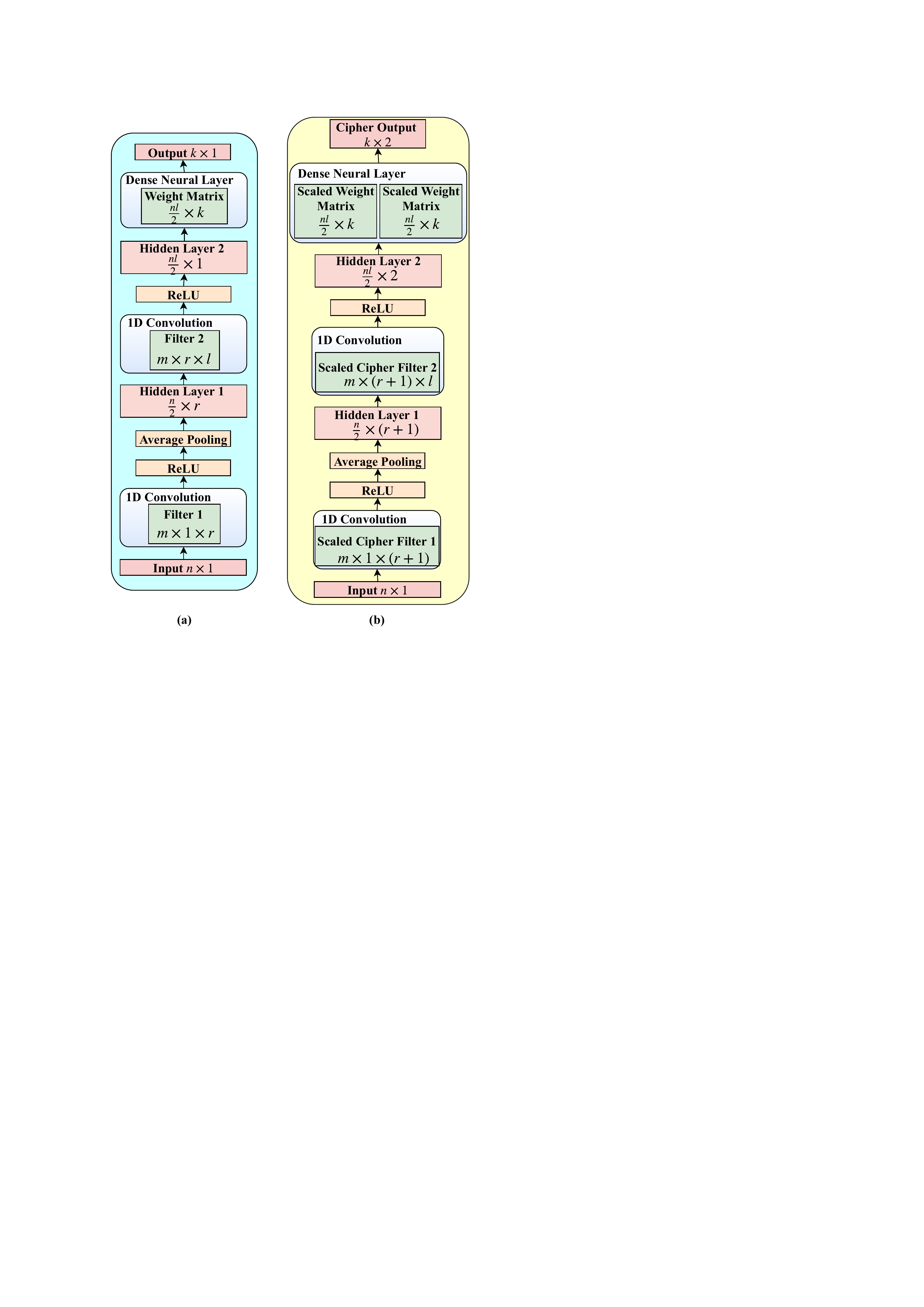}
\caption{Illustration of our matrix-pair-wise encryption strategy: (a) a CNN-based machine learning model in plainspace and (b) the corresponding CNN-based machine learning model in ciperspace.}\label{Fig:Method-2}
\end{figure}

Our matrix-pair-wise encryption strategy is performed on a pair of neighboring weight matrices. To illustrate our strategy, we use a CNN as an example shown in Fig.~\ref{Fig:Method-2}. As illustrated in Fig.~\ref{Fig:Method-2}, this CNN has two convolution layers having the dimensions of $m \times 1 \times r$ and $m \times r\times l$, respectively, where $r$ is the number of convolution filters in the first convolutional layer and $l$ is the number of convolution filters in the second convolutional layer.

In our encryption strategy, the dimension of encryption is carefully selected such that there is no dimension mismatch while executing the CNN-based machine learning model in cipherspace. Additionally, the first convolution layer is encrypted via IVHE scheme by leveraging the 3rd dimension, which results in a scaled and encrypted convolutional filter with dimension $m \times 1 \times (r+1)$. Similarly, the second convolution layer is encrypted on the 2nd dimension, which results in a scaled and encrypted convolutional filter with dimension $m \times (r+1) \times l $. By doing so, the convolution operation can be performed as a pair without any dimension mismatch.

\subsection{Learning-Model Fusion Mechanism}\label{sec:fuse}

As illustrated in Fig.~\ref{Fig:DL}, the success of our decentralized and secure computing paradigm requires an efficient learning-model fusion mechanism, with which the application initiator is able to integrate the verified private models and achieve an effective MetaModal for the targeted data-driven task. For designing the learning-model fusion mechanism, it is necessary to treat the computing models, provided by the computing contributors, as separate entities, to ensure the fused structure is dynamic.

\begin{figure*}[]
\center
\includegraphics[width=0.6\textwidth]{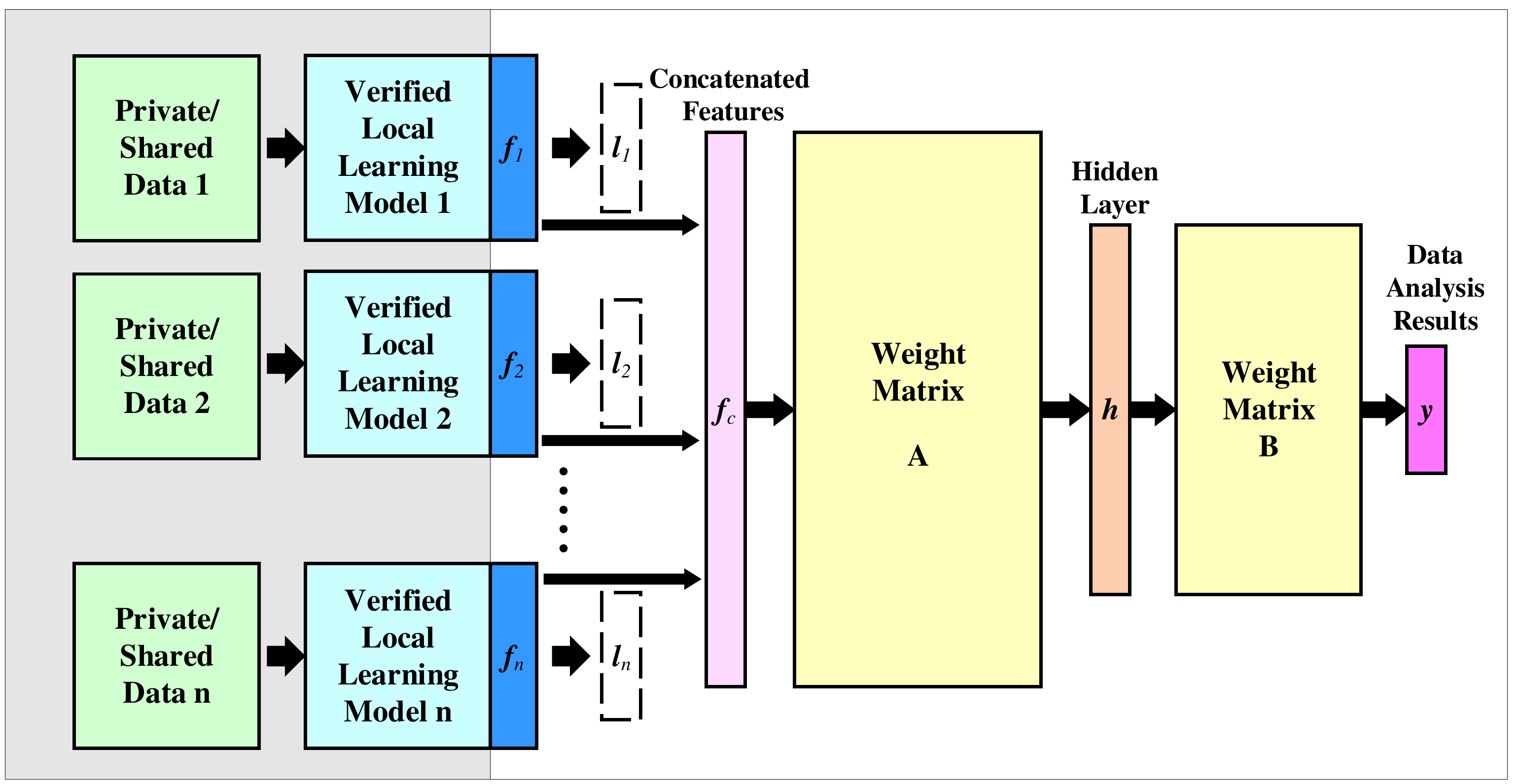}
\caption{\label{Fig:ModDrop} Illustration of our proposed learning-model fusion mechanism.}
\center
\end{figure*}

Figure~\ref{Fig:ModDrop} illustrates the structure of our proposed fusion mechanism. In our mechanism, the upper-layer feature vectors $\bf{f}_i$ from individual verified private models are concatenated to form a concatenated feature layer $\bf{f}_c$. As shown in~\ref{Fig:ModDrop}, the upper-layer feature vector $\bf{f}_i$ can be the input of the output layer for the private model $i$. One fully-connected neural network (NN) is designed to fuse the features characterized by the individual private models. This fully-connected NN uses the concatenated feature layer $\bf{f}_c$ as its input layer and has a hidden layer $\bf{h}$ with the length of $\sum_{i=1}^{n} \mid l_{i} \mid$, where $\mid l_{i} \mid $ is the number of the labeled classes in the $i$th private model. TO design the fully-connected NN, it is essential to design its weight matrices $\bf A$ and $\bf B$. Currently, we consider two strategies for learning the optimum values for weight matrices $\bf A$ and $\bf B$ and designing the learning-model fusion mechanism.

\subsubsection{Strategy I}\label{sec:s1}
In this strategy, the weight matrices $\bf A$ and $\bf B$ are initialized randomly without any additional constraints and are optimized via the backpropagation algorithm~\cite{Goodfellow-et-al-2016} in which the $i$th element in the hidden layer $\bf h$ and the $j$th element in the output layer $\bf y$ are calculated as follows: \begin{equation}\label{eq:hidden_basic}
 h_j = \sum_{i=1}^{\mid {\bf fc} \mid}{A_{ij}^T} \cdot fc_{i}
\end{equation}
\begin{equation}\label{eq:label_basic}
y_j = \dfrac{exp(\sum_{i=1}^{\mid {\bf h} \mid} {B_{ij}^T} \cdot h_{i})}{\sum_{k=1}^{d} exp(\sum_{i=1}^{\mid {\bf h} \mid} B_{ik}^T \cdot h_{i})}
\end{equation}

\subsubsection{Strategy II}\label{sec:s2}

This strategy is developed to achieve to goals: (1)learning the uniqueness of the features characterized by the individual verified private models, and (2) exploring the correlation amongst the features presented by the individual private models. To achieve this goal, a gradual fusion is designed, in which the weight matrix $\bf{A}$ is initialized as a concatenated matrix formulated in Eq.~(\ref{eq:init_A}) and the weight matrix $\bf{B}$ is initialized with the concatenation of the identity matrices, as formulated in Eq.~(\ref{eq:init_B}).
\begin{equation}\label{eq:init_A}
{\bf{A}}_{init} = \begin{bmatrix}
    \bf{W_{1}} & \bf{0} & \bf{0} & \dots  & \bf{0} \\
    \bf{0} & \bf{W_{2}} & \bf{0} & \dots  & \bf{0} \\
     \bf{0} & \bf{0} & \bf{W_{3}}  & \dots  & \bf{0} \\
    \vdots & \vdots & \vdots & \ddots & \vdots \\
    \bf{0} & \bf{0} & \bf{0} & \dots  & \bf{W_{n}}
\end{bmatrix}
\end{equation} where a diagonal weight matrix $\bf{W}_i$, which has the dimension of $({\mid {\bf{f}}_i \mid} , {\mid {\bf{l}}_i \mid})$, is initialized randomly.

\begin{equation}\label{eq:init_B}
\footnotesize
{\bf{B}}_{init} = \begin{bmatrix}
    w_{111}=1 & 0 & 0 & \dots  & 0 \\
    0 & w_{122}=1 & 0 & \dots  & 0 \\
    0 & 0 & w_{133}=1  & \dots  & 0 \\
    \vdots & \vdots & \vdots & \ddots & \vdots \\
    0 & 0 & 0 & \dots  & w_{1dd}=1 \\
    w_{211}=1 & 0 & 0 & \dots  & 0 \\
    0 & w_{222}=1 & 0 & \dots  & 0 \\
    0 & 0 &  w_{233}=1  & \dots  & 0 \\
    \vdots & \vdots & \vdots & \ddots & \vdots \\
    0 & 0 & 0 & \dots  & w_{2dd}=1 \\
    \vdots & \vdots & \vdots & \ddots & \vdots \\
    w_{n11}=1 & 0 & 0 & \dots  & 0 \\
    0 & w_{n22}=1 & 0 & \dots  & 0 \\
    0 & 0 & w_{n33}=1  & \dots  & 0 \\
    \vdots & \vdots & \vdots & \ddots & \vdots \\
    0 & 0 & 0 & \dots  & w_{ndd}=1 \\
\end{bmatrix}
\end{equation} where $n$ is the number of the verified private models to be fused and $d$ denotes the number of class labels.

The elements of the weight matrices $\bf{A}$ and $\bf{B}$ are optimized by using our gradual fusion method that consists of two stages. In the initial stage, only the diagonal non-zeros weights of matrix $\bf{A}$ by using backpropagation algorithm, which targets at learning the uniqueness of the features characterized by the individual private models. In the second stage, all of the weights in $\bf{A}$ are updated by using backpropagation algorithm, which targets at exploring the correlations between the features characterized by the individual private models. Accordingly, the $i$th element in the hidden layer $\bf h$ and the $j$th element in the output layer $\bf y$ are calculated as follows:

\begin{equation}\label{eq:hidden_updated}
 h_j = \sum_{i=1, A_{ij} \in \bar {\bf W}}^{\mid {\bf fc} \mid}{A_{ij}^T} \cdot fc_{i} + \gamma \sum_{i=1, A_{ij} \notin \bar {\bf W}}^{\mid {\bf fc} \mid}{A_{ij}^T} \cdot fc_{i}
\end{equation}where the parameter $\gamma$ is set as $0$ in the intial stage and as $1$ in the final stage, {\footnotesize{ $\bar {\bf W} = {\bf W_p} \gets  \sum_{k=1}^{p-1}\mid {\bf l_k} \mid < i \leq \sum_{k=1}^{p}\mid {\bf l_k} \mid  $}}.
\begin{equation}\label{eq:label_updated}
\footnotesize
y_j = \dfrac{exp(\sum_{i=1,B_{ij}=w_{jpp}}^{\mid {\bf h} \mid} {B_{ij}^T} \cdot h_{i}+ \gamma \sum_{i=1,B_{ij} \neq w_{jpp}}^{\mid {\bf h} \mid} {B_{ij}^T} \cdot h_{i}))}{\sum_{k=1}^{d} exp(\sum_{i=1, B_{ik} = w_{ipp}}^{\mid {\bf h} \mid} B_{ik}^T \cdot h_{i} + \gamma \sum_{i=1,  B_{ik} \neq w_{ipp}}^{\mid {\bf h} \mid} B_{ik}^T\cdot h_{i})}
\end{equation} where $p \in \lbrace 1,2,3 \dots ,d \rbrace$, and $d$ is the number of labeled classes.

\section{Simulation Results}\label{sec:sim}

In this section, the performance of our proposed blockchain-powered decentralized and secure computing paradigm is evaluated by considering three case studies. To achieve this goal, we develop a Blockchain-powered Software-defined networking (SDN)-based testbed as detailed in Section~\ref{testbed}.

\subsection{Blockchain-Powered SDN-Based Testbed}~\label{testbed}

\begin{figure*}[!htb]
\center
\includegraphics[width=0.9\textwidth]{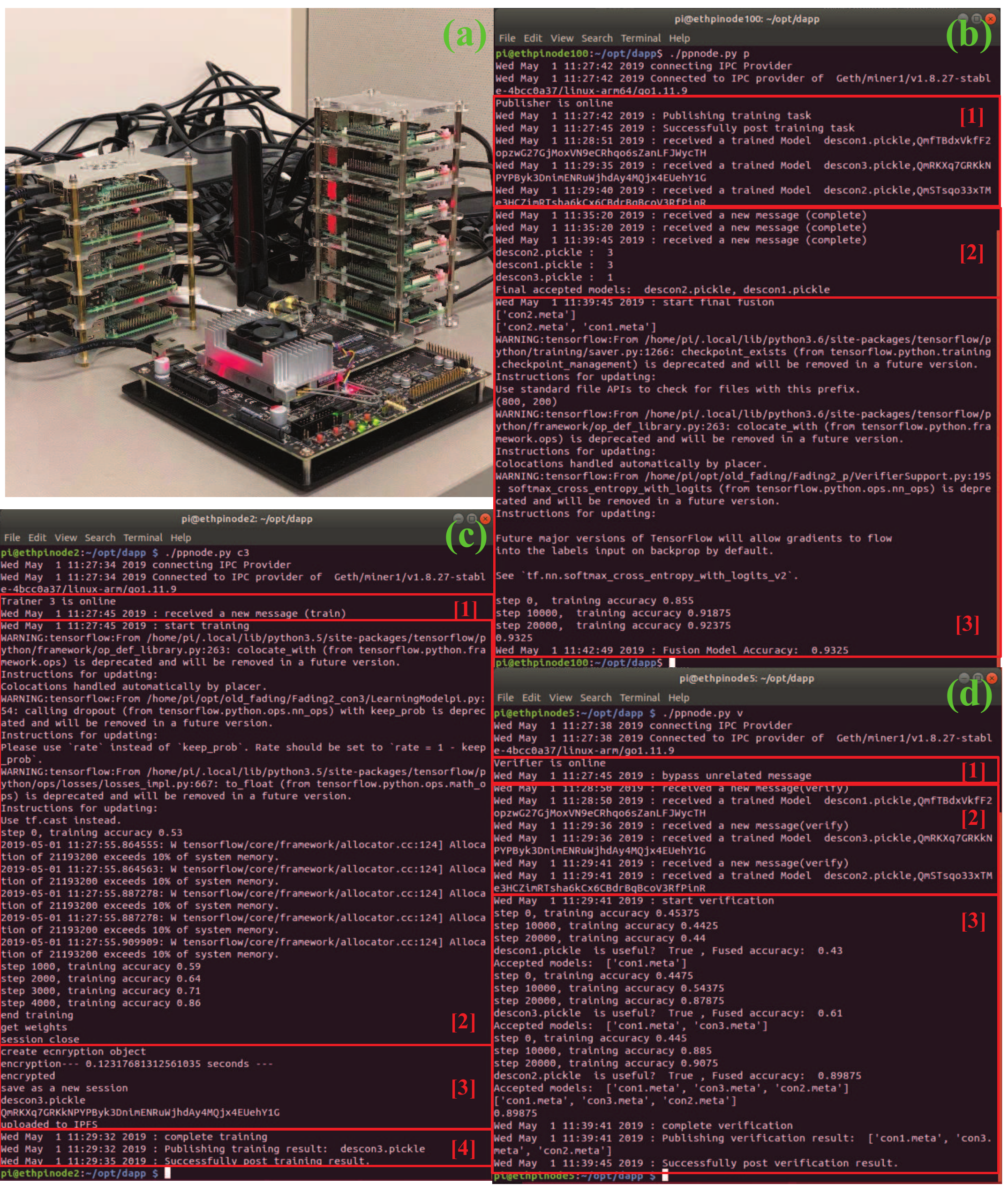}
\caption{(a) A picture of the testbed, and the terminals of (b) Application Initiator, (c) Computing Contributor, and (d) Verification Contributor.}\label{fig:Pi}
\end{figure*}

\begin{figure*}[!htb]
\center
\includegraphics[width=0.7\textwidth]{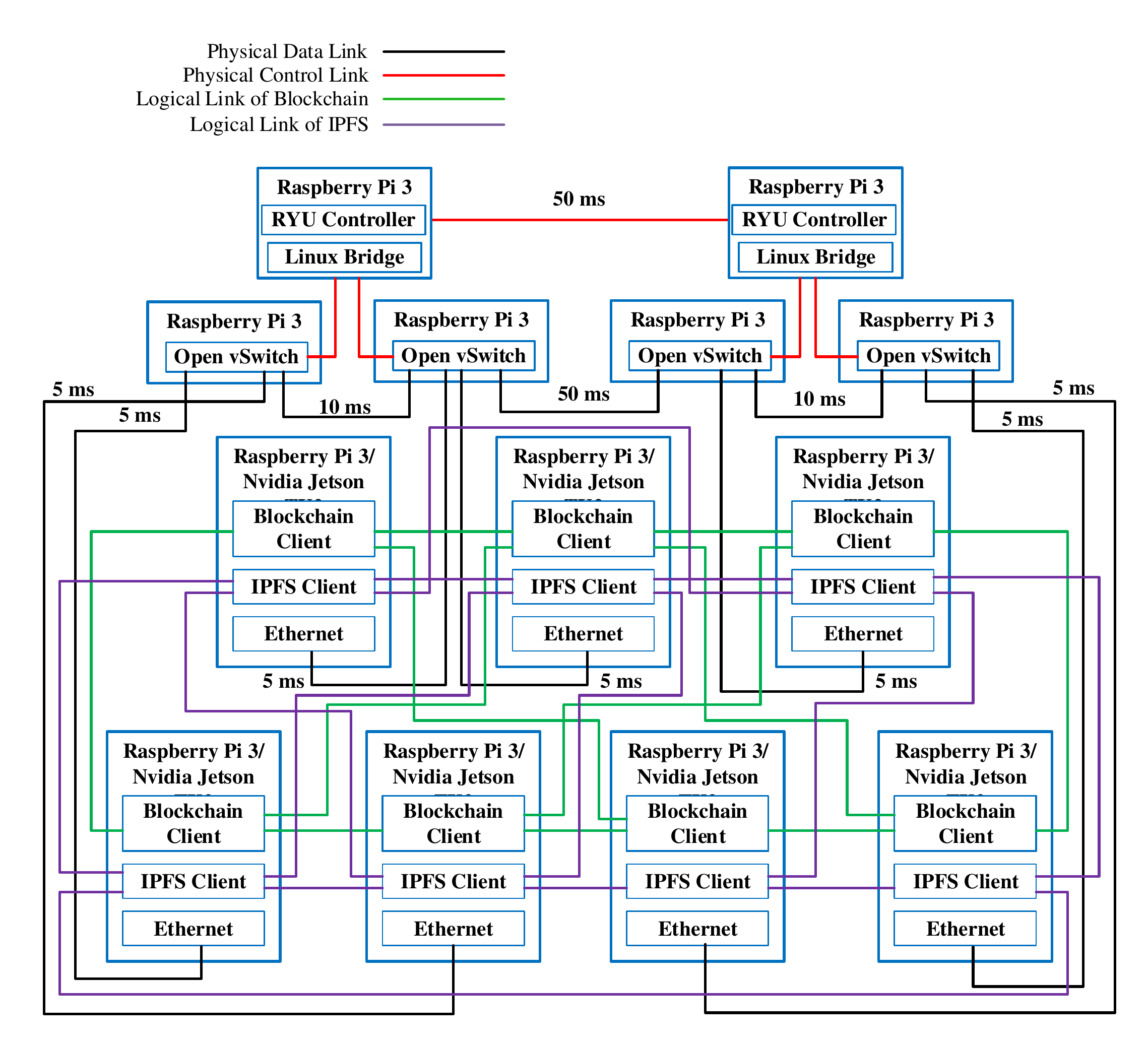}
\caption{(a) Illustration of implementing our blockchain-powered and SDN-based decentralized and secure computing testbed.}\label{Fig:BlockchainNetwork}
\end{figure*}

Fig.~\ref{fig:Pi}~(a) shows a picture of our Blockchain-powered SDN-based testbed, in which the computing nodes, including one application initiators, three computing contributors, and three verification contributors, are simulated by using two popular embedded systems, Raspberry PI and NVIDIA Jetson. The communications and cooperative computing amongst the computing nodes are supported by the blockchain middleware and the SDN-based peer-to-peer networking layer. The integration of the computing layer, blockchain middleware, and the SDN-enabled networking layer in our computing testbed are illustrated in Fig.~\ref{Fig:BlockchainNetwork}. Each computing node, which is simulated via Raspberry PI or NVIDIA Jetson, has (1) one or multiple blockchain clients to interact with the blockchain middleware, (2) multiple ethernet ports to interact with the SDN-enabled networking layer, and (3) one or more decentralized storage (DS) clients to enable the decentralized storage systems.

In our testbed, the blockchain middleware is developed by exploiting a decentralized application (DApp) on the Ethereum platform in which the engine and language of smart contract were Ethereum Virtual Machine (EVM) and Solidity. The consensus protocol is set to be Proof-of-Authority (PoA). The block interval is set to $15$~s and all sealers are initialized with a small number of tokens. Additionally, the smart contract in Ethereum is written by Solidity and the blockchain event listener client is leveraged to provide the interface of the logging facilities of the Ethereum virtual machine. Every computing node, which is willing to participate in a certain decentralized computing task via the blockchain smart contract, is required to send a transaction through Ethereum blockchain to contract with a considerate amount of security deposit of tokens. By doing so, the smart contract is able to record the identities of all the participants and forward the tasks such as training and verification to the participants appropriately. The deployment of smart contract is considered as a transaction as well, which provides the other nodes with the address necessary to access the smart contract. As shown in Fig~\ref{Fig:BlockchainNetwork}, the DS system in our testbed is realized by using Interplanetary File System (IPFS) that utilizes the peer-to-peer network to share and store hypermedia on the Internet. Since the whole simulation was running on a private network, a shared swarm key was used to authorize reliable nodes to join IPFS. Furthermore, our computing layer is developed in Python because most frameworks of machine learning are realized there. Web3.py and ipfs-api.py provide access to Ethereum blockchain and IPFS clients to realize the transactions and file sharing, respectively. Tensorflow is a deep learning framework with Python bindings, which is adopted for the task of machine learning in our testbed.

To further demonstrate our testbed, the screenshot of the terminals for a computing initiator, a computing contributor, and a verification contributor during one experiment are shown in Figs.~\ref{fig:Pi}(b)-(d), respectively. As shown in Fig.~\ref{fig:Pi}(b), the application initiator's operation comprises three main processes that are identified with three red rectangular boxes: (1) being initialized to publish one data-driven task, characterize the objectives and constraints of the task via smart contract, upload the data to be shared to IPFS, and share the public key generated via homomorphic encryption interface, (2) receiving the verified models that are concluded according to the majority votes of all the active verification contributors, and (3) implementing decryption and fusion on the received verified models. As shown in Fig.~\ref{fig:Pi}(c), the computing contributor's operation mainly consists of four processes: (1) participating the data-driven task, (2) terminating the training task when the accuracy meets the given/self-defined criterion, (3) encrypting the achieved private model via homomorphic encryption interface and uploading the encrypted private model to IPFS, and (4) announcing the encrypted private model for assessment. As shown in Fig.~\ref{fig:Pi}(d), the verification contributor's operation includes three main processes: (1) being passively and randomly selected and initialized for the given data-driven task, (2) receiving the all announced private models, and (3) verifying the received private models according to the criterion defined in smart contract and publishing the verification results.

\subsection{Case Study I}\label{sec:cs1}
\begin{figure}[!htb]
\center
\includegraphics[width=0.45\textwidth]{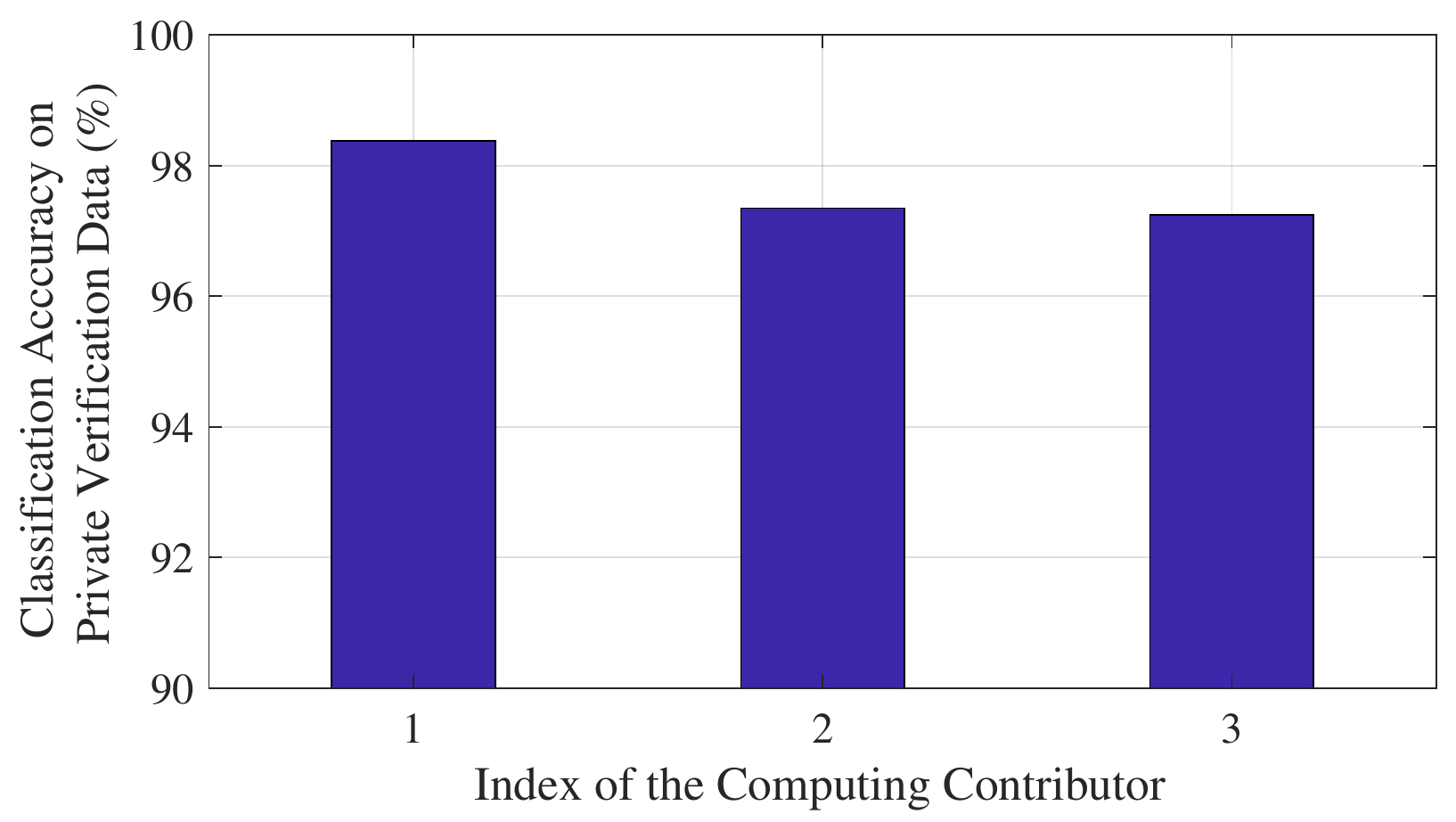}
\caption{\label{Fig:prov_accu_s1} Accuracy obtained by each computing contributor by using their local verification data.}
\center
\end{figure}

\begin{figure}[!htb]
\center
\includegraphics[width=0.48\textwidth]{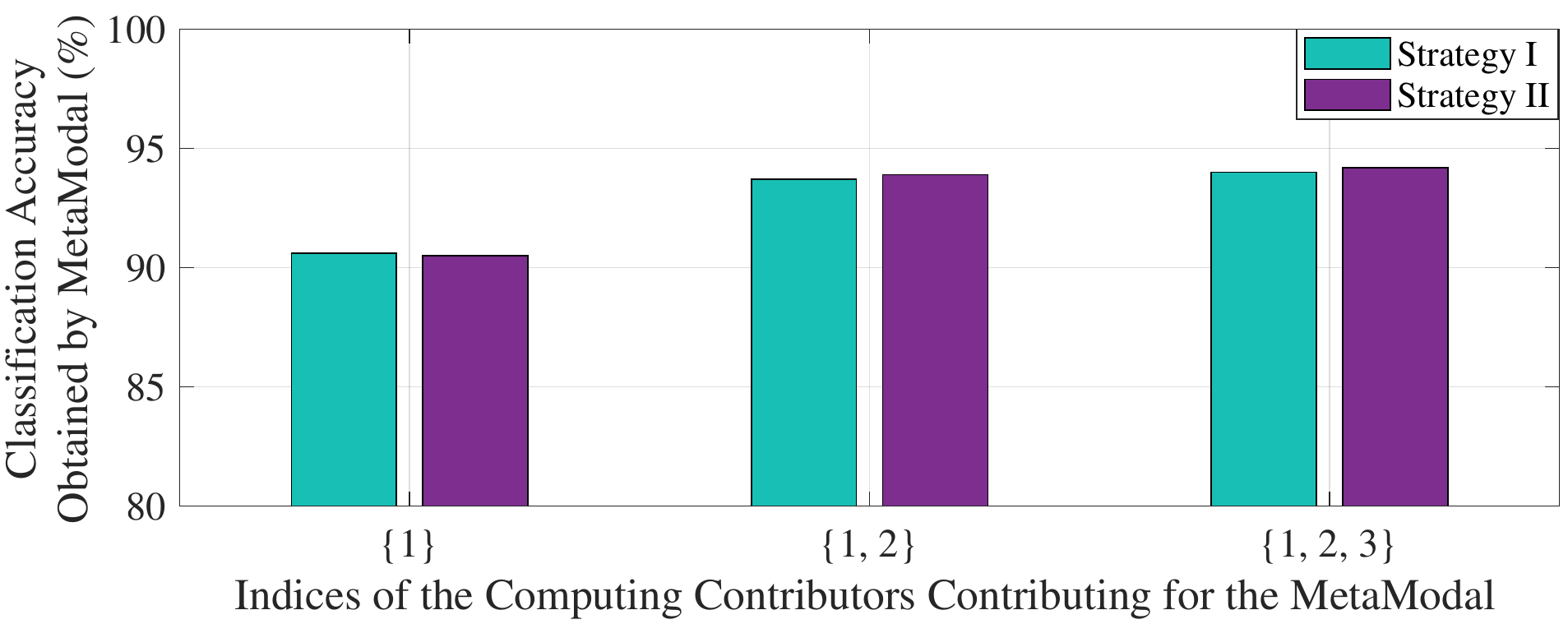}
\caption{\label{Fig:fused_accu_compare} Comparison of average classification accuracy obtained by MetaModal using strategies I and II versus the indices of the computing contributors contributing to the MetalModal.}
\center
\end{figure}

In this case study, we focus on evaluating the performance of our proposed blockchain-powered decentralized and secure computing paradigm in a secure environment. In other words, the functionality of the homomorphic encryption interface is not considered. We assume one application initiator publishes a data-driven task on classifying the $10$ image classes provided by the MNIST handwritten digit database~\cite{lecun1998mnist} and three computing contributors, and three verification contributors, which are considered to be randomly selected, participate in the task. We consider that each computing contributor develops its own CNN-based machine learning model whose structure parameters are summarized in Table~\ref{Tab:CNN}. As shown in Table~\ref{tab:data_dist}, the local data available to each computing contributor only present a partial view of the whole dataset. The accuracy of the CNN models developed by the individual computing contributors for classifying their local verification data is shown in Fig.~\ref{Fig:prov_accu_s1}, all of which are above $90$~\%. In this scenario, the criterion of determining whether the local learning model is trained successfully is achieving the accuracy of $90$~\%. Therefore, all of the three private CNN models are encrypted and published to the three verification contributors for assessment. After receiving the verified private models according to the majority voting amongst the verification contributors, the application initiator fuses the models by using the two strategies introduced in Section~\ref{sec:fuse} to achieve the MetaModal. Assume that the application initiator decrypts and fuses the private models as soon as receiving them, the classification accuracy achieved by the MetaModal is shown in Fig.~\ref{Fig:fused_accu_compare}. From Fig.~\ref{Fig:fused_accu_compare}, it is clear that the classification accuracy increases as more private models are fused. This is reasonable since the individual private models are achieved using the local data that only characterize the partial features of the whole dataset. Furthermore, it can also see from Fig.~\ref{Fig:fused_accu_compare} that Fusion Strategy~II slightly outperforms the Fusion Strategy~I when fusing multiple private models.

\begin{table*}[!htb]
\centering
\caption{Parameters of the local learning models considered in Case Study I}
\label{Tab:CNN}
\begin{tabular}{cccccc}
\hline
\multicolumn{4}{c}{CNN parameters} \\ \hline
 & Computing  & Computing  & Computing \\
 & Contributor 1 & Contributor 2 & Contributor 3 \\ \hline
Inputs & \multicolumn{3}{c}{28x28 images} \\ \hline
Convolution layer l & 32 5x5 kernels & 64 5x5 kernels & 32 5x5 kernels \\ \hline
Pooling layer 1 & \begin{tabular}[c]{@{}c@{}}2x2 \\ maximum pooling\end{tabular} & \begin{tabular}[c]{@{}c@{}}2x2 \\ maximum pooling\end{tabular} & \begin{tabular}[c]{@{}c@{}}2x2 \\ maximum pooling\end{tabular}  \\ \hline
Convolution layer 2 & 16 5x5 kernels & 32 5x5 kernels & 32 10x10 kernels \\ \hline
Pooling layer 2 & \begin{tabular}[c]{@{}c@{}}2x2 \\ maximum pooling\end{tabular} & \begin{tabular}[c]{@{}c@{}}2x2 \\ maximum pooling\end{tabular} & \begin{tabular}[c]{@{}c@{}}2x2 \\ maximum pooling\end{tabular} \\ \hline
Convolution layer 3 & 8 2x2 kernels & 16 2x2 kernels & 16 4x4 kernels \\ \hline
\begin{tabular}[c]{@{}c@{}}Reshaped vector \\ (Convolution layer 3 output \\ are flatten as a vector of size)\end{tabular} & 7x7x8=392 & 7x7x16=784 & 7x7x16=784 \\ \hline
hidden layer & \multicolumn{3}{c}{Fully connected hidden layer  with size 500} \\ \hline
Output & \multicolumn{3}{c}{10 labels with softmax activation} \\ \hline
Training method & \multicolumn{3}{c}{Adam Optimizer} \\
Batch size & \multicolumn{3}{c}{50} \\
Learning rate & \multicolumn{3}{c}{0.0001} \\
Maximum number of epochs & \multicolumn{3}{c}{100} \\ \hline
\end{tabular}
\end{table*}

\begin{table}[!htb]
\centering
\caption{Summary of the dataset available to the individual contributors in Case Study I }
\label{tab:data_dist}
\scriptsize{
\begin{tabular}{|c|c|c|c|}
\hline
Contributor & \begin{tabular}[c]{@{}c@{}}Set of \\ Labels\end{tabular} & \begin{tabular}[c]{@{}c@{}}No. \\ Training \\ Data\end{tabular} & \begin{tabular}[c]{@{}c@{}}No.\\  Verification \\ Data\end{tabular} \\ \hline
Verifiers & \begin{tabular}[c]{@{}c@{}}\{0, 1, 2, 3, 4,\\5, 6, 7, 8, 9\}\end{tabular} & 1000  & 1000 \\ \hline
\begin{tabular}[c]{@{}c@{}}Computing \\ Contributor 1\end{tabular} & \{0, 1, 2, 3, 4\} & 1000  & 1000 \\ \hline
\begin{tabular}[c]{@{}c@{}}Computing \\ Contributor 2\end{tabular} & \{0, 6, 7, 8, 9\} & 1000  & 1000 \\ \hline
\begin{tabular}[c]{@{}c@{}}Computing \\ Contributor 3\end{tabular} & \{5, 6, 7, 8, 9\} & 1000  & 1000 \\ \hline
\end{tabular}}
\end{table}

\subsection{Case Study II}\label{sec:cs2}
In this case study, we focus on evaluating the performance of homomorphic encryption interface in our proposed decentralized and secure computing paradigm by using the testbed shown in Figs.~\ref{fig:Pi}~(a)~and~\ref{Fig:BlockchainNetwork}. The local data that can be accessed by the individual computing contributors are detailed in Table~\ref{tab:data_dist_sc2}, which show that the individual computing contributors only have partial view of the entire dataset. Additionally, we assume that each computing contributor locally trains a CNN private learning model with a similar structure as illustrated in Table~\ref{Tab:CNN_sc2}. To perform our homomorphic encryption interface, the floating-point input data and weight parameters of the CNN-based learning models are converted to be integer by introducing appropriate scaling factors $p$ and $q$ as shown in Fig~\ref{Fig:Method-1}. In this simulation, we evaluate the performance of the final MetaModal when the computing contributors select the scaling factors $p = 2^5$ or $2^7$ and $q = 1000$. Furthermore, we assume that in the encryption interface of the computing contributors, the private CNN-based learning models are encrypted via the first convolution layers. In the encryption interface of the application initiators, the verified private models are decrypted and fused to achieve MetaModals via Fusion Strategy II. The accuracy of the MetaModal achieved with and without homomorphic encryption interface are shown in Table~\ref{tab:result_sc2}, respectively. From Table~\ref{tab:result_sc2}, we can see the MetaModal achieved by using the encryption interface with the scaling factor $p=2^7$ outperforms that with the scaling factor $p=2^5$ and achieves comparable accuracy as the original MetaModal. This is reasonable because that the errors caused by rounding the parameters of the CNN-based private models to integers increases when the scaling factor is lower.

\begin{table}[]
\centering
\caption{Summary of the local data available to the individual contributors in Case Study II}
\label{tab:data_dist_sc2}
\scriptsize{
\begin{tabular}{|c|c|c|c|}
\hline
Contributor & \begin{tabular}[c]{@{}c@{}}Set of \\ Labels\end{tabular} & \begin{tabular}[c]{@{}c@{}}No. \\ Training \\ Data\end{tabular}  & \begin{tabular}[c]{@{}c@{}}No.\\  Verification \\ Data\end{tabular} \\ \hline
\begin{tabular}[c]{@{}c@{}}Computing \\ Contributor 1\end{tabular} & \{0, 1, 2, 3, 4, 5, 6\} & 1000  & 1000 \\ \hline
\begin{tabular}[c]{@{}c@{}}Computing \\ Contributor 2\end{tabular} & \{0, 1, 2, 3, 4, 8, 9\} & 1000  & 1000 \\ \hline
\begin{tabular}[c]{@{}c@{}}Computing \\ Contributor 3\end{tabular} & \{0, 1, 2, 6, 7, 8, 9\} & 1000  & 1000 \\ \hline
\end{tabular}}
\end{table}

\begin{table}[]
\centering
\caption{Parameters of the CNN-based private models in Case Study II}
\label{Tab:CNN_sc2}
\begin{tabular}{cccccc}
\hline
\multicolumn{2}{c}{CNN parameters} \\ \hline
Inputs & \multicolumn{1}{c}{28x28 images} \\ \hline
Convolution layer l & 32 5x5 kernels  \\ \hline
Pooling layer 1 & \begin{tabular}[c]{@{}c@{}}2x2 \\ average pooling\end{tabular}  \\ \hline
Convolution layer 2 & 16 5x5 kernels \\ \hline
Pooling layer 2 & \begin{tabular}[c]{@{}c@{}}2x2 \\ average pooling\end{tabular}  \\ \hline
\begin{tabular}[c]{@{}c@{}}Reshaped vector \\ (Pooling layer 2 output \\ are flatten as a vector of size)\end{tabular} &  7x7x16=784 \\ \hline
Output & \multicolumn{1}{c}{10 labels with softmax activation} \\ \hline
Training method & \multicolumn{1}{c}{Adam Optimizer} \\
Batch size & \multicolumn{1}{c}{50} \\
Learning rate & \multicolumn{1}{c}{0.0001} \\
Maximum number of epochs & \multicolumn{1}{c}{100} \\ \hline
\end{tabular}
\end{table}

\begin{table}[]
\centering
\caption{Comparison of The Accuracy Achieved by MetaModals in Different Situations in Case Study II}
\label{tab:result_sc2}
\begin{tabular}{|c|c|c|c|}
\hline
\multirow{2}{*}{\begin{tabular}[c]{@{}c@{}}Indices of \\ Fused Models\end{tabular}} & \multicolumn{3}{c|}{Classification Accuracy for Fused Models (\%)} \\ \cline{2-4}
 & Original & \begin{tabular}[c]{@{}c@{}}Encrypted \\ ($p = 2^7$)\end{tabular} & \begin{tabular}[c]{@{}c@{}}Encrypted \\ ($p = 2^5$)\end{tabular} \\ \hline
\{1\} & 91.5 & 91.5 & 89.6 \\ \hline
\{1, 2\}& 94.8 & 94.7 & 93.0 \\ \hline
\{1, 2, 3\} & 95.8 & 95.8 & 93.8 \\ \hline
\end{tabular}
\end{table}

\subsection{Case Study III}\label{sec:cs4}

In this case study, we compare our proposed computing paradigm to a widely recognized distributed machine learning paradigm, federated learning~\cite{mcmahan2016communication} in a data-driven task on classifying the $10$ image classes provided by the MNIST handwritten digit database~\cite{lecun1998mnist}. In federated learning, the local contributors initialize their model by referring the global initialized weights and then update their own models for a certain numbers of local epochs. The weights of the local models are averaged as the new global weights, which completes the first round of model update. The updated global weights are distributed to local computation contributors for a next round of model updating procedure. This process is called Federated Averaging (FedAvg).

To achieve a fair comparison, we assume the decentralized computing is conducted in a secure environment. Therefore, in our computing paradigm, the functionality of our homomorphic encryption interface is not considered and only one verification contributor is needed. For federated learning, we assume that the global average process ignores the random selection of local weights and collect all the weights achieved by the available local models.  In addition, we assume that each computing contributor adopts a three-layer CNN model in both of the computing paradigms. In our computing paradigm, there are three computing contributors, each of which conducts up to $100$ training epoches locally, one verification contributor, and one application initiator. In federated learning, there are four local contributors training the local models, each of which involves in five rounds of model updating and conducts $20$  training epoches locally in each round.

Furthermore, in this case study, we consider two scenarios. In the first scenario, we consider a diverse set of local training samples of the MNIST data available to the local contributors. Each of these local training samples includes the data associated with all of the $10$ image class labels, and the distribution of the local training samples are not identical amongst these contributors. In the second scenario, we consider that data are distributed amongst contributors in a way that only part of the $10$ image classes are available to most of the contributors. The data distribution used in this scenario is similar to Table~\ref{tab:data_dist}.

\begin{figure}[]
\center
\includegraphics[width=0.45\textwidth]{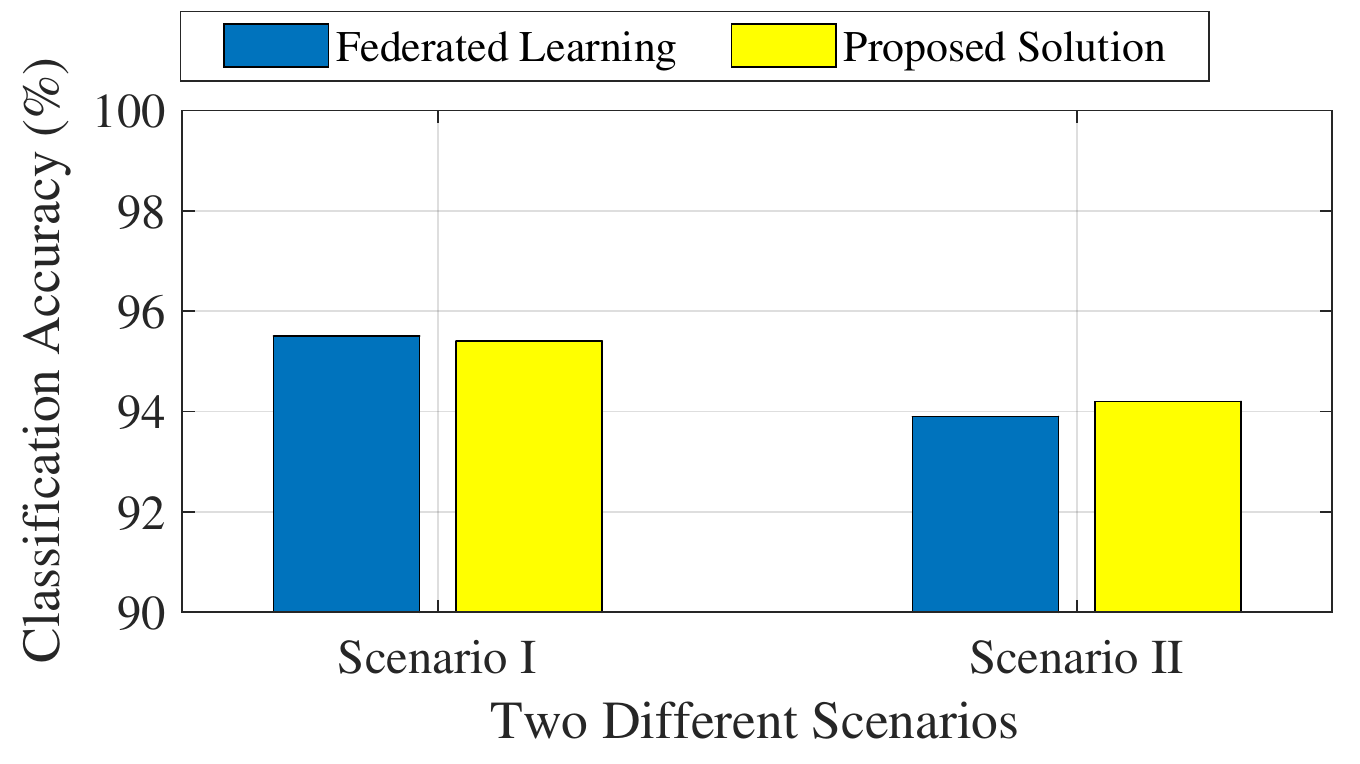}
\caption{\label{Fig:fed_compare} Performance comparison between our computing paradigm and federated learning in two different scenarios}
\center
\end{figure}

The comparison result is shown in Fig.~\ref{Fig:fed_compare}, from which it can be seen that both computing paradigms achieve a comparable classification accuracy. In the first scenario where all the contributors have a good view of the entire dataset, federated learning slightly outperforms our framework. In the second scenario, where most contributors only have partial views of the entire dataset, our computing solution is slightly better than the federated learning. Furthermore, our computing solution provides higher integrity security via removing the authoritative controlling agent and introducing the verification contributors. The main functionality of verification contributors is for security purpose rather than for increasing the final accuracy, which also results in the relatively lower accuracy achieved by our solution in the first scenario shown in Fig.~\ref{Fig:fed_compare}. Furthermore, the communication cost required by federated learning is higher than that required by our solution. This is caused by the centralized management and the repeated collection of the local weights during the process FedAvg.

\subsection{Case Study IV}\label{sec:cs3}

In this case study, we evaluate the performance of our decentralized and secure computing paradigm with the functionality of the homomorphic encryption interface in a data-driven task of detecting a fading channel by using link power data. The confidential data used in this task are from Link Power database provided by NASA Glenn Research Center. In the simulation, we assume that one application initiator publishes this data-driven task and there are three computing contributors and three randomly selected verification contributors participating in the task. The link power data are randomly divided into $4$ sets each of which contains $10000$ training data and $1000$ testing data. Of these four sets, three sets are used as the local data, each of which is available to one of the three computing contributors. The other set is considered accessible by  assigned to the three randomly selected verification contributors in the testbed. Additionally, we assume that each computing contributor participating in the task adopts a CNN model with $2$ convolution layers and ReLU activation function. In this case study we consider both Element-wise and Matrix-Pair-wise homomorphic encryption strategies.

To demonstrate the operation of our homomorphic encryption interface, Table~\ref{tab:con_scIV} summarizes the accuracies obtained by each computing contributor with its private model and local testing data with and without the encryption interface. From these results, it is clear that implementing the encryption interface slighted reduces the classification accuracies, which is the cost to achieve higher security and privacy preservation. Additionally, Table~\ref{tab:ver_scIV} shows the classification accuracy achieved by the MetaModal of the application initiator, which is obtained by fusing the individual verified private model via fusion strategy II. From the results in Table~\ref{tab:ver_scIV}, which is obtained considering the functionality of homomorphic encryption interface, it is clear that although the final result is satisfactory the encryption interface slightly reduces the accuracy due to the introduction of randomization as explained in Section~\ref{sec:intvect1} and rounding error as discussed in Section~\ref{sec:cs2}.

Furthermore, we study the impact of the homomorphic encryption interface on the execution time. Tables~\ref{tab:timing} summarizes the overhead for encryption and the time consumed to achieve a local private model by each computing contributor with and without encryption interface. Homomorphic encryption interface is based on Element-wise Encryption. The results are obtained on a Corei5 CPU for 1000 testing data. ML models are executed on TensorFlow graphs. Encryptions and decryptions are performed using Python Numpy API. It can be seen that the execution time with encryption interface is only $2$ times higher compared with that without encryption interface.

\begin{table*}[]
\center
\caption{The Accuracy of Fading Channel Detection Accuracy Using The Private CNN-Based Learning Models}
\label{tab:con_scIV}
\begin{tabular}{|c|c|c|c|}
\hline
\multirow{2}{*}{Model} & \multicolumn{3}{c|}{Detection Accuracy (\%)} \\ \cline{2-4}
 & Without Encryption &  Element-wise Encryption & Matrix-Pair-wise Encryption \\ \hline
Computing Contributor 1 & 93 & 91.6 & 91.7 \\ \hline
Computing Contributor 2 & 93.1 & 92.4 & 92.0 \\ \hline
Computing Contributor 3 & 93.2 & 92.3 & 92.1 \\ \hline
\end{tabular}
\end{table*}

\begin{table*}[]
\center
\caption{Detection Accuracy Achieved by using MetalModal}
\label{tab:ver_scIV}
\begin{tabular}{|c|c|c|}
\hline
\multirow{2}{*}{Model} & \multicolumn{2}{c|}{Detection Accuracy (\%)} \\ \cline{2-3}
& Element-wise Encryption & Matrix-Pair-wise Encryption \\ \hline

\{1, 2\} & 94.5 & 94.5 \\ \hline
\{1, 2, 3\} & 95.5 & 95.0 \\ \hline
\end{tabular}
\end{table*}

\begin{table*}[]
\center
\caption{The Impact of the Element-wise Encryption Interface on Execution Time}
\label{tab:timing}
\begin{tabular}{|c|c|c|c|}
\hline
\multirow{2}{*}{Model} & \multicolumn{3}{c|}{Execution Time (ms)} \\ \cline{2-4}
 & \begin{tabular}[c]{@{}c@{}}Overhead for \\ Encryption\end{tabular} & \begin{tabular}[c]{@{}c@{}}Execution without \\ Encryption\end{tabular} & \begin{tabular}[c]{@{}c@{}}Execution with \\ Encryption\end{tabular} \\ \hline
\begin{tabular}[c]{@{}c@{}}Computing \\ Contributor 1\end{tabular} & 7.7 & 60 & 126 \\ \hline
\begin{tabular}[c]{@{}c@{}}Computing \\ Contributor 2\end{tabular} & 7.6 & 57 & 123 \\ \hline
\begin{tabular}[c]{@{}c@{}}Computing \\ Contributor 3\end{tabular} & 7.6 & 59 & 125 \\ \hline
\end{tabular}
\end{table*}

\section{Conclusions}\label{sec:con}

Availability of computation power and data are two of the main reasons for the success of machine learning in a variety of application areas. However, both acquisition of processing power and data can be expensive. In many instances these challenges are addressed by relying on an outsourced cloud-computing vendor. However, although these commercial cloud vendors provide valuable platforms for data analytics, they can suffer from a lack of transparency, security, and privacy-preservation. Furthermore, reliance on cloud servers prevents applying big data analytics in environments where the computing power is scattered. Therefore, more effective computing paradigms are required to process private and/or scattered data in suitable decentralized ways for machine learning. To pave the way to achieve this goal, a decentralize, secure, and privacy-preserving computing paradigm is proposed in this paper to enable an asynchronized cooperative computing process amongst scattered and untrustworthy computing nodes that may have limited computing power and computing intelligence. This paradigm is designed by exploring blockchain, decentralized learning, homomorphic encryption, and software defined networking(SDN) techniques. The performance of the proposed paradigm is evaluated by considering different scenarios and comparing to a widely recognized distributed machine learning paradigm, federated learning, in the simulation section.

%

\ifCLASSOPTIONcompsoc
  \section*{Acknowledgments}
\else
  \section*{Acknowledgment}
\fi
This research work was supported by NASA under Grant 80NSSC17K0530. The authors would like to thank Praveen Fernando for assistance with the preliminary work.
\bibliographystyle{IEEEtran}%
\bibliography{review,gihan,moeinref}

\end{document}